\documentclass[acmsmall]{acmart}

\usepackage[normalem]{ulem}
\usepackage{epstopdf,tikz}
\usepackage{hyperref}
\usepackage[most]{tcolorbox}
\usepackage{hyperxmp}
\usepackage{enumitem}
\usepackage{pifont,epsfig, endnotes}
\usepackage{amsmath}
\usepackage{graphicx, float, array}
\usepackage{color,textcomp,xcolor}
\usepackage{booktabs}
\usepackage{algpseudocode,algorithm}
\usepackage{subfloat}
\usepackage{wrapfig}
\usepackage{subcaption}
\usepackage{enumerate}
\usepackage{makecell}
\usepackage{balance}
\usepackage{filecontents}
\usepackage{xspace}
\usepackage{caption}
\usepackage{libertine,environ}
\usepackage{tikz}

\newcommand{\fcirc}[1]{%
  \tikz[baseline=(char.base)]{
    \node[shape=circle, fill=black, inner sep=1.2pt] (char) {\color{white}\scriptsize #1};
  }%
}
\newcommand{\sys}{\MakeLowercase{d}-HNSW\xspace}

\newcommand{\ra}[1]{\textcolor[rgb]{0,0,0}{#1}}
\newcommand{\rb}[1]{\textcolor[rgb]{0,0,0}{#1}}
\newcommand{\rc}[1]{\textcolor[rgb]{0,0,0}{#1}}
\newcommand{\rd}[1]{\textcolor[rgb]{0,0,0}{#1}}
\newcommand{\re}[1]{\textcolor[rgb]{0,0,0}{#1}}

\renewcommand\footnotetextcopyrightpermission[1]{} 
\acmSubmissionID{2}
\makeatletter
\let\@authorsaddresses\@empty
\makeatother

\AtBeginDocument{%
  \providecommand\BibTeX{{%
    \normalfont B\kern-0.5em{\scshape i\kern-0.25em b}\kern-0.8em\TeX}}}

\begin{document}

\title{d-HNSW: A High-performance Vector Search Engine on Disaggregated Memory}

\author{Fei Fang}
\email{ffang6@ucsc.edu}
\author{Yi Liu}
\email{yliu634@ucsc.edu}
\author{Chen Qian}
\email{cqian12@ucsc.edu}
\affiliation{
    \institution{University of California, Santa Cruz}
    \city{Santa Cruz}
    \state{California}
    \country{United States}}

\begin{abstract}
Efficient vector search is essential for powering large-scale AI applications, such as LLMs. \re{Existing solutions are designed for monolithic architectures where compute and memory are tightly coupled. Recently, disaggregated architecture breaks this coupling by separating compution and memory resources into independently scalable pools to improve utilization. However, applying vector database on disaggregated memory system brings unique challenges to system design due to its graph-based index.} We present \sys, the first RDMA-based vector search engine optimized for disaggregated memory systems. \sys preserves HNSW’s high accuracy while addressing the new system-level challenges introduced by disaggregation: 1) network inefficiency from pointer-chasing traversals, 2) non-contiguous remote memory layout induced by dynamic insertions, 3) redundant data transfers in batch workloads, and 4) resource underutilization due to sequential execution.
\sys tackles these challenges through a set of hardware-algorithm co-designed techniques, including 1) balanced clustering with a lightweight representative index to reduce network round-trips and ensure predictable latency, 2) an RDMA-friendly graph layout that preserves data contiguity under dynamic insertions, 3) query-aware data loading to eliminate redundant fetches across batch queries, and 4) a pipelined execution model that overlaps RDMA transfers with computation to hide network latency and improve throughput. Our evaluation results in a public cloud show that \sys achieves up to $< 10^{-2}\times$ query latency and $>100\times$ query throughput compared to other baselines, while maintaining a high recall of 94\%.
\end{abstract}

\ccsdesc[500]{Networks~Storage area networks}

\keywords{Vector similarity search; HNSW; Distributed graph computing}

\maketitle


%
%
%
%
%

\section{Introduction}
\label{sec:intro}
Similarity search on vectors~\cite{wang2021milvus,wang2024vector,hm-ann,zhang2019grip,rummy,auncel,chen2018sptag,jiang3,isca} aims to identify the most similar vectors from a large dataset given a query vector. 
Instead of exact nearest neighbor search in vector data, approximate nearest neighbor (ANN) algorithms 
can accelerate search by orders of magnitude while sacrificing a little accuracy~\cite{aumuller2020ann, jiang1,jiang2,jiang3,isca, hpca, micro}. 
Vector databases leverage ANN search techniques to support efficient retrieval, making them a core building block for a wide range of machine learning applications~\cite{faiss,liu2024retrievalattention} such as recommendation systems~\cite{wentao} and retrieval-augmented generation (RAG)~\cite{ragcache, jiang4,jiang5} for large language models (LLMs)~\cite{liu2024retrievalattention}. For example, they power semantic search engines by indexing text and multimedia embeddings for relevance ranking~\cite{pan2024survey}. In recommendation systems~\cite{wentao}, items and users are embedded into high-dimensional vectors so that nearest neighbor search finds similar products or content for personalization. 
In RAG systems~\cite{ragcache}, a vector database retrieves semantically relevant documents based on the user prompt’s embedding, allowing LLMs to generate responses with external knowledge rather than being limited to the information encoded in model parameters. As these AI-driven applications~\cite{wentao, memserve} continue to grow in scale and complexity, the demand for high-performance, scalable vector databases has exploded, transforming them from a useful tool into a critical part of the AI systems stack for the next generation of AI.

Disaggregated memory systems~\cite{wang2023disaggregated, lee2021mind,wang2022case} represent a significant shift from traditional monolithic architectures by decoupling memory and computation into separate, scalable pools. In this design, compute pools consist of nodes equipped with rich CPU resources, while memory pools consist of nodes configured with abundant DRAM and storage. This paradigm is enabled by high-performance network fabrics like Remote Direct Memory Access (RDMA)~\cite{guideline}. RDMA provides high throughput (40-800 Gbps) and microsecond-level latency, allowing compute nodes to access memory on remote machines directly, bypassing the remote CPU to reduce overhead and improve efficiency. Leveraging RDMA, the disaggregated architecture has been successfully applied to a wide range of data-intensive systems, including file systems~\cite{dlsm,wang2022sherman}, key-value stores~\cite{shen2023fusee,wei2021xstore,zuo2022race,li2023rolex,outback}, and transactional databases~\cite{zhang2022ford,zhang2024motor}. For example, recent industry efforts, such as DeepSeek's 3FS~\cite{3fs,deepseek}, have demonstrated RDMA's benefits for demanding AI workloads. The unique challenges posed by the complex data structures and access patterns of vector search have hindered the adoption of disaggregated designs.

\begin{wrapfigure}[7]{r}{0.45\textwidth}
  \centering
  \vspace{-5ex}
  \includegraphics[width=0.4\textwidth]{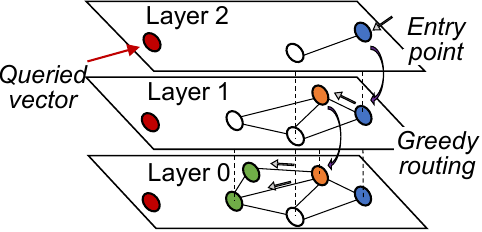}
  \vspace{-1.8ex}
  \caption{Graph-based vector search index: HNSW.}
  \label{fig:hnsw}
\end{wrapfigure}

Modern vector datasets exhibit heterogeneous and dynamic resource demands. For example, compute-intensive query workloads often fluctuate independently of storage demands as dimensions scale. Disaggregation ~\cite{tsai2020disaggregating,wang2023disaggregated} is \textit{uniquely suited} to address this imbalance by enabling independent scaling of memory pools and compute instances. \textbf{However, no existing design has applied RDMA-based disaggregated memory to vector databases}. \textbf{We argue that this architectural gap presents a vital, unexplored opportunity.} 

To understand why this gap exists, we examine the internal mechanics of the most prominent ANN algorithms~\cite{kdtree,lsh,wang2021comprehensive,wang2024vector}. Among the various similarity search algorithms, graph-based approaches~\cite{nsg,hnsw} have consistently demonstrated state-of-the-art performance, offering a superior trade-off between accuracy and query latency. Hierarchical Navigable Small World (HNSW)~\cite{hnsw} has emerged as a widely adopted graph-based index that balances vector search accuracy and efficiency. HNSW constructs a multi-layered graph where nodes are vectors and edges represent proximity. This hierarchical structure enables a greedy search that is both fast and accurate: long-range edges in upper layers allow for rapid traversal across the vector space, while shorter edges in lower layers facilitate precise, fine-grained navigation.

In this work, we introduce \sys, a fast, RDMA-based vector similarity search engine designed for disaggregated memory system. \sys aims to bridge the gap between high-performance vector similarity search and the emerging disaggregated architecture in datacenters, ensuring scalability and efficiency in handling high-throughput data queries. 
The disaggregated memory pool provides abundant memory resources, allowing us to store both the HNSW index and all original floating-point vector data on it. The disaggregated compute instances handle data requests and reply on one-sided RDMA primitives to directly access the index and vectors, bypassing the memory instances' CPUs. \textbf{\sys addresses the following four challenges that have not been resolved in the existing systems.}

\rc{The first challenge stems from adapting HNSW's greedy search~\cite{lighttraffic} to a networked environment, specifically regarding the issue of \textit{"pointer-chasing"}: each memory access depends on the result of the previous one, because the next distance computation candidate will be decided from neighbors of vectors in the last rounds. The algorithm navigates the index by traversing an unpredictable search path, where each expansion compares the query vector against a set of candidate nodes, including the current node and its neighbors. In a disaggregated system, fetching each vector along this path would trigger an independent, high-latency network round-trip, making the process prohibitively slow. To mitigate this excessive pointer-chasing over the network, we partition the dataset into smaller clusters.} A lightweight \textit{representative index} is used to identify which clusters are most likely to contain the top-$k$ nearest neighbors, and only these targeted partitions need to be read from remote memory. 
It is crucial that this process yields \textit{balanced clusters}, ensuring each partition contains a similar number of vectors. 
Balanced partitions provide predictable performance and uniform network cost.

Second, 
when new vectors are inserted, they need to be stored in available memory while ensuring fast index access for different partitions. To maximize memory efficiency, remote data are typically stored in a dense, contiguous layout. However, this design leaves little reserved capacity for in-place insertions. In disaggregated memory systems, relocating large volumes of remote data incurs high computational and network costs, forcing newly inserted vectors to be appended to a global memory region. This append-only design scatters vectors from the same partition across non-contiguous memory locations, requiring multiple RDMA\_READ during query processing and increasing latency. To address this, we propose an \textit{RDMA-friendly graph index layout} that enables efficient data queries while supporting dynamic vector insertions.

Third, processing queries independently~\cite{faiss} often leads to repeated data requests, as different queries may require the same data partition. Given the limited DRAM cache in the compute pool, avoiding redundant data transfers becomes particularly important. 
To address this, we introduce \textit{query-aware data loading}, a technique that analyzes the requirements of an entire batch, eliminates repeated data fetching, and ensures each required partition is fetched from the memory pool \textbf{only once}. This design significantly reduces network traffic and improves overall query throughput.

The fourth challenge is overcoming the inefficiency of sequential execution of operations in disaggregated memory, where valuable CPU resources remain idle during I/O-intensive network operations. 
We design a \textit{pipelined execution model} that orchestrates network transfers and CPU computations to run in parallel. Specifically, \sys overlaps the RDMA transfer of one sub-HNSW data partition with the CPU-intensive search computation of a previously fetched partition. This approach reduces exposed latency by overlapping data movement with computation, enabling high utilization of both network and CPU resources. By carefully managing this pipeline, \sys significantly reduces query latency and improves overall system throughput.

\rb{The design of \sys is distinct from the existing distributed vector search engines~\cite{wang2021milvus, DBLP:journals/corr/abs-2112-09924,wang2021milvus}. Unlike disaggregated architectures, these systems couple computation resources and data together tightly, relying on host-centric execution and RPC-based access.} As a result, they cannot efficiently handle the above four challenges. 
In contrast, \sys is a hardware-algorithm co-design, aligning its search algorithms and data structures directly with RDMA fabric. Our solution enables efficient graph-based vector searches in disaggregated memory. Evaluation results show that \sys achieves $< 10^{-2}\times$ query latency and up to $>100\times$ query throughput compared to other state-of-the-art vector search baselines, while preserving high accuracy. In summary, we make the following contributions:
\begin{itemize}[leftmargin=4ex,itemsep=0.ex,parsep=0.ex]
  \vspace{-1.ex}
  \item We present \sys, to the best of our knowledge, the first RDMA-based vector similarity search engine designed for disaggregated memory.
  \item We propose a suite of co-designed techniques to overcome the limitations of running graph-based search over a network. These include (a) balanced data partitioning to ensure predictable performance, (b) an RDMA-friendly memory layout for efficient dynamic updates, (c) query-aware data loading to minimize network traffic, and (d) a pipelined execution model to hide network latency.
  \item We implement \sys as a prototype running in a public cloud platform~\cite{cloudlab}. The evaluation shows that \sys reduces query latency and improves query throughput by more than two orders of magnitude. 
  An anonymous repository is available~\cite{artifact} for review purposes.
  \vspace{-1.ex}
\end{itemize}


\section{Background, Opportunities, and Challenges}
\label{sec:background}
\vspace{-.5ex}
This section first provides essential background on graph-based vector search with HNSW and the architecture of RDMA-based disaggregated memory. We then detail the core challenges of implementing HNSW in this architecture, which motivate the design of \sys.

\vspace{-1.5ex}
\subsection{Vector similarity search}
\label{sec:background:vectordb}

Vector similarity search is crucial for retrieving high-dimensional data in modern ML applications, such as RAG~\cite{ragcache} for LLMs. While traditional methods like KD-trees~\cite{kdtree} and LSH~\cite{lsh} struggle with the curse of dimensionality, graph-based ANN search algorithms~\cite{nsg,hnsw} become the dominant approach, offering superior performance. These methods construct a navigable graph where nodes represent data points and edges represent their proximity, enabling fast query traversal.

Among all ANN search algorithms, Hierarchical Navigable Small World (HNSW)~\cite{hnsw} is a widely adopted algorithm that balances search accuracy and efficiency. HNSW constructs a multi-layered graph where each layer is a subset of the one below it, as illustrated in Fig.~\ref{fig:hnsw}. The number of vectors in each layer decreases exponentially towards the top, with the uppermost layer ($L_{max}$) serving as a coarse-grained ``express lane'' for long-range navigation, while the bottom layer ($L_0$) contains all vectors in the dataset for fine-grained search refinement. HNSW builds a hierarchical, multi-layer graph by assigning new vectors to randomly determined maximum layers following an exponentially decaying probability distribution. Vectors are inserted by greedy routing from the top layer down to the appropriate level, establishing bidirectional links with their nearest neighbors at each layer. The density of connections, controlled by the degree of graph connectivity, and the dynamic candidate list size,
impacts accuracy, memory usage, and construction speed, creating a fundamental trade-off in graph-based ANN indexing.

In HNSW, a query search starts at a global entry point in the top layer and greedily traverses the graph to find the local optimum, which becomes the entry point for the next lower layer. At each layer, the algorithm maintains a candidate list of promising nodes during graph traversal, with its size controlled by the parameter $e$. This layer-by-layer descent continues, with the search becoming progressively finer-grained. Upon reaching the bottom layer ($L_0$), the search continues until the candidate list converges, after which the final top-$k$ nearest neighbors are returned. A higher $e$ typically improves \textit{recall} (the proportion of correctly returned data records out of all top-$k$ records in the dataset) by exploring more nodes but increases query latency.By leveraging this multi-layered, small-world graph structure, HNSW avoids the need for exhaustive comparisons and significantly prunes the search space, making it one of the most effective and widely-used ANN algorithms in modern vector databases~\cite{ChenW18, WangL12, WangWZTGL12, WangWJLZZH14, faiss, wang2021milvus, hnsw}.


\vspace{-2ex}
\subsection{RDMA-based disaggregated memory}
\label{sec:background:rdma}

Modern datacenters increasingly adopt disaggregated memory architectures to overcome the scaling limitations of traditional monolithic servers~\cite{tsai2020disaggregating,wang2023disaggregated, jiang1,jiang3,hpca}. In these architectures, compute and memory resources are decoupled into independent pools, enabling memory to scale without adding underutilized CPUs and compute to scale independently of memory, leading to flexible hardware scalability, high resource utilization, and efficient data sharing. Compute nodes, typically equipped with abundant CPU resources but limited local DRAM, access a large pool of remote memory hosted on dedicated memory nodes via a high-performance network fabric. \ra{Prior work has explored KV stores~\cite{li2023rolex,tsai2020disaggregating,zuo2022race,zuo2021one} and tree-based indexes~\cite{dlsm,wang2024optimizing,lu2024dex,wang2022sherman} over RDMA-based disaggregated memory. These systems typically issue one-shot or bounded remote accesses, yielding a small and predictable set of RDMA reads per request, which aligns well with one-sided RDMA execution.}

RDMA technologies (e.g., RoCE~\cite{roce}, Infiniband~\cite{guideline}) enable reliable and in-order packet delivery, making them well-suited for indexing structures in disaggregated memory systems. RDMA bypasses the kernel network stack and allows user-space applications to directly read from or write to remote memory. 
Different from two-sided RDMA systems~\cite{kalia2016fasst,outback} that require remote CPU involvement, 
one-sided RDMA systems~\cite{li2023rolex,luo2023smart,mitchell2016balancing,wang2022sherman,zuo2022race} avoid remote CPU intervention, making them more suitable for building low-latency, high-throughput systems in disaggregated architectures by reducing additional latency and CPU overhead on memory nodes.


\vspace{-1.5ex}
\subsection{Opportunity and Challenges}
\label{sec:background:challenges}

\begin{figure*}[t]
    \begin{minipage}[t]{0.32\textwidth}
        \centering
        \vspace{-18.8ex}
        \includegraphics[width=\linewidth]{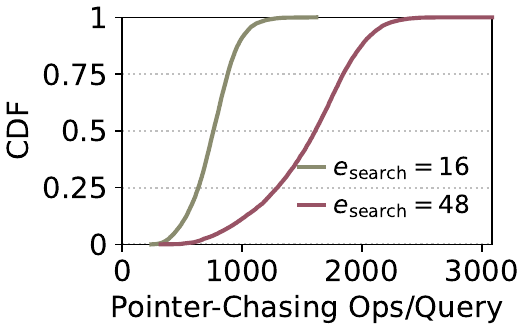}
        \vspace{-6ex}
        \caption{\re{CDF of HNSW pointer-chasing operations per query for different search candidate list size on SIFT1M.}}
        \label{fig:cha1}
    \end{minipage}
    \hspace{-1ex}
    \begin{minipage}[t]{0.65\textwidth}
        \centering
        \begin{subfigure}[t]{0.49\linewidth}
            \centering
             \vspace{-18.5ex}
             \includegraphics[width=\linewidth]{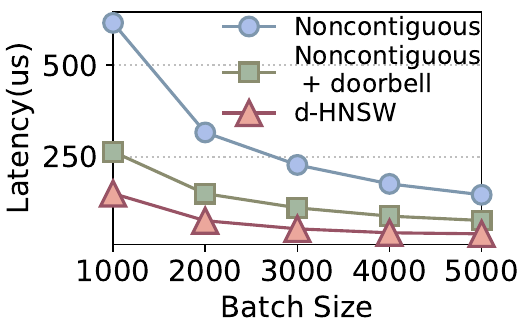}
            \vspace{-4ex}
            \captionsetup{margin={1.2em,0em}}
            \caption{SIFT1M}
            \label{fig:cha2:sift1M}
        \end{subfigure}
        \hspace{-1.5ex}
        \begin{subfigure}[t]{0.495\linewidth}
            \centering
            \includegraphics[width=\linewidth]{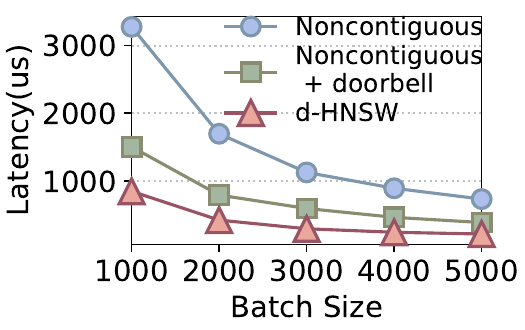}
            \vspace{-4ex}
            \captionsetup{margin={1.5em,0em}}
            \caption{GIST1M}
            \label{fig:cha2:gist1M}
        \end{subfigure}
        \vspace{-3.ex}
        \caption{\re{Search latency after 10K insertions.}}
        \vspace{-1.ex}
        \label{fig:cha2}
    \end{minipage}
    \vspace{-5ex}
\end{figure*}


\textbf{Opportunity.}
Traditional monolithic server architectures present a significant challenge for deploying large-scale vector databases. These systems impose fixed CPU and memory resources, leading to inefficient resource allocation. For instance, a workload that is memory-intensive but computationally light (e.g., serving a large, static index) would leave expensive CPU cores idle. Conversely, a compute-heavy workload (e.g., batch indexing or complex queries) on a smaller dataset would waste memory capacity. This tight coupling results in inefficient resource utilization, with either compute or memory resources often left idle. Disaggregated systems address this imbalance by decoupling compute and memory into independent, scalable pools. Then, resources can be provisioned on demand to match workload requirements. Expensive memory can be pooled and shared across multiple compute instances, while CPU resources can be dynamically allocated to handle fluctuating query loads. This elasticity not only improves hardware utilization and reduces operational costs but also enables more efficient and scalable vector database deployments.

\begin{wrapfigure}[9]{r}{0.34\textwidth}
  \hfill
  \vspace{-3.ex}
  \includegraphics[width=\linewidth]{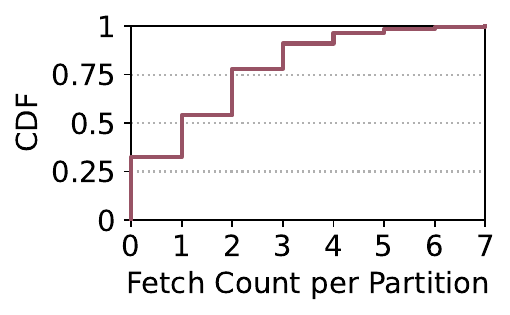}
  \vspace{-6.5ex}
  \captionsetup{margin={0.5cm,0cm}} 
  \caption{\re{CDF of fetch count per partition in one batch.}}
  \label{fig:cha3}
  \captionsetup{margin=0cm}
\end{wrapfigure}

\textbf{Challenge 1: Network Latency incurred by Pointer-Chasing in Graph.} 
\label{sec:motivation:1}
\re{Algorithms like HNSW use greedy search with unpredictable traversal paths~\cite{lighttraffic}, requiring comparisons at each step. Consequently, HNSW exhibits pronounced \emph{pointer-chasing}. The traversal repeatedly follows data-dependent links to neighbors unknown in advance, and resolves that neighbor’s vector before it can perform the next round of distance comparisons and decide subsequent hops. This induces an irregular access pattern that is difficult to prefetch effectively. The number of such unpredictable pointer-chasing operations varies substantially across queries and search candidate size. Fig.~\ref{fig:cha1} illustrates the magnitude and variability of this overhead: the CDF of pointer-chasing operations is routinely on the order of \emph{hundreds to thousands} of operations, with a heavy tail approaching $\sim$3{,}000 ops/query. We use the number of distance computations as a proxy for pointer chasing, since each comparison typically requires retrieving a neighbor vector from memory. Moreover, these neighbor vectors are scattered across memory and accessed in a data-dependent order, offering limited spatial locality opportunity for effective prefetching. In disaggregated systems, these dependent accesses translate into many \textit{network round-trips}, substantially slowing down search.}
\re{Partitioning the dataset can reduce these round-trips by restricting the search to a small subset of clusters, but keeping partitions balanced while preserving search quality is hard. State-of-the-art partitioning strategies like k-means~\cite{lloyd1982least} and KAHIP~\cite{sandersschulz2013} often create uneven cluster sizes, leading to unpredictable performance: queries to large partitions are slow, while others are fast. Also, even if similar vectors were assigned to different partitions, recall will drop because relevant candidates could be missed during searching.} This causes high tail latency and low accuracy. To address this, we use \textit{a lightweight representative index} to find promising clusters for each query and apply balanced clustering that preserves vector similarity within partitions while ensuring uniform partition sizes.
\begin{figure}[!t]
	\centering
	\includegraphics[scale=0.42]{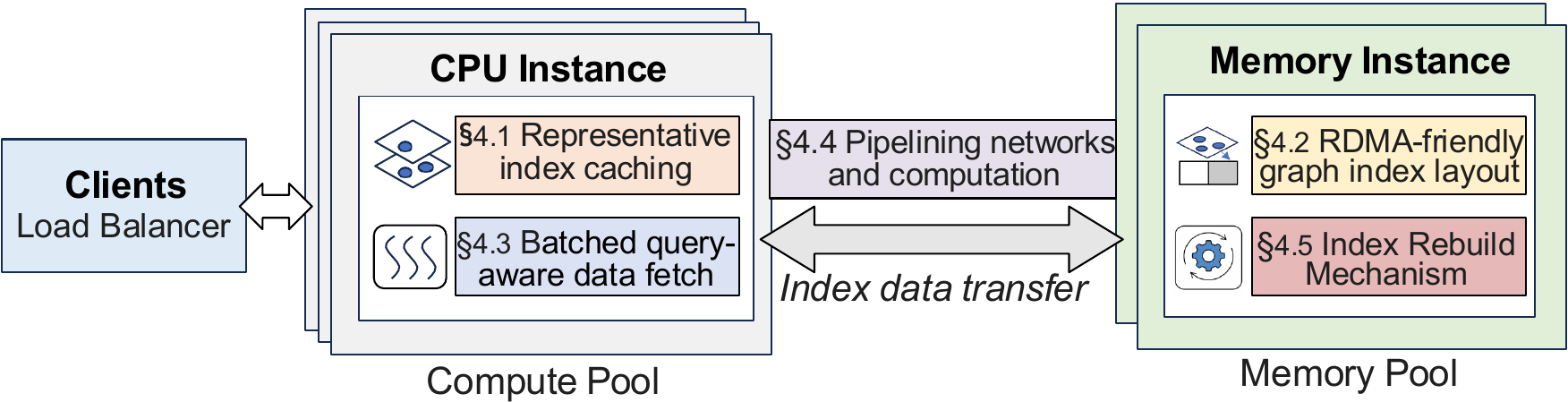}
    \vspace{-1.5ex}
	\caption{Overview of \sys.}
	\label{fig:overview}
    \vspace{-6.1ex}
\end{figure}

\textbf{Challenge 2: Static Data Layout under Dynamic Updates.} The second challenge involves organizing data in remote memory to support both efficient access and dynamic vector insertions. When new vectors are inserted into the system, they must be stored in available memory while maintaining efficient access patterns for existing partitions. Na\"ive allocation strategies place newly inserted vectors into a global append-only space, causing data belonging to the same partition to become scattered across non-contiguous remote memory regions.  As a result, query processing must issue multiple RDMA\_READ operations to retrieve data that logically belongs to a single partition, each incurring network latency and PCIe overhead. This effect becomes more severe as the database scales and undergoes frequent updates. \re{To isolate the impact of memory fragmentation under dynamic updates, we run a microbenchmark that measures query latency after inserting 10,000 vectors into a SIFT1M and GIST1M. We compare our RDMA-friendly layout (\sys) with two non-contiguous baselines that append newly inserted vectors outside the original memory region and record the partition they belong to. During search, this non-contiguous memory design requires multiple RDMA\_READs to fetch both the original partition and the appended updates for the same partition, increasing effective RTTs. We additionally evaluate a doorbell-batched variant that posts these RDMA\_READ together to reduce RTT. We vary the query batch size (1k--5k) and report average per-query latency. As shown in Fig.~\ref{fig:cha2}, doorbell batching reduces the latency of fragmented layouts by roughly 2$\times$, but \sys consistently outperforms both non-contiguous baselines by 1.8-2.1$\times$ across all batch sizes and datasets, demonstrating the importance of RDMA-friendly data layout under dynamic inserts.} Designing data layouts that are aware of NIC-level architectural constraints is essential for efficient RDMA usage. So we tackle this challenge by designing \textit{an RDMA-friendly graph index layout} that enables efficient data queries while supporting dynamic vector insertions through careful memory organization. As a result, an entire partition can later be fetched via one RDMA\_READ even after many updates.

\textbf{Challenge 3: Redundant Data Fetching in Batch Processing.} Batch processing often leads to redundant data fetching when multiple queries need the same partitions. \re{Fig.~\ref{fig:cha3} illustrates the frequency with which each partition is fetched across queries within a batch, showing that a substantial fraction of partitions are repeatedly accessed (fetch count $\geq 2$) by multiple queries in the same batch. As a result, without coordination, the system may fetch the same data repeatedly, wasting network bandwidth and overloading the limited DRAM cache. As batch sizes grow and queries show spatial locality, this redundancy causes cache thrashing, with frequently used partitions repeatedly evicted and re-fetched, degrading performance.} To address this, we use \textit{query-aware data loading} that analyzes the entire batch, eliminates repeated fetches, and fetches each partition only once.

\textbf{Challenge 4: Resource Underutilization During Sequential Execution.} \rd{A naive sequential approach alternates between data fetching and computation, leaving CPUs idle during transfers and wasting network bandwidth during computation. This underutilization is costly in disaggregated systems, where network transfer times can be significant. The challenge lies in reducing stalls caused by data dependencies, ensuring that computation can proceed without waiting for remote data. We address this by using a \textit{pipelined execution model} that overlaps RDMA transfers with CPU search, hiding network latency behind computation.}

In summary, these four challenges highlight the need for specialized techniques that can effectively bridge the gap between traditional vector search algorithms and the constraints of disaggregated memory system architectures. \ra{Motivated by this gap, \sys bridges this gap by co-designing the index and execution under RDMA constraints, combining a representative index with balanced clustering that restricts traversal to a limited subset, RDMA-friendly layouts maintain near-contiguous index under insertions, query-aware loading to avoid redundant transfers, and pipelined execution overlaps RDMA transfers with CPU search. Together, these techniques convert irregular vector-graph access into predictable RDMA operations.}
\vspace{-1.5ex}
\section{\sys Design Overview}
\label{sec:design:overview}

 \rd{We present the design of \sys, a high-performance vector similarity search engine built for a disaggregated memory environment, which includes four main techniques. \sys exploits the characteristics of RDMA-based memory data accessing and graph-based index HNSW to realize fast vector query processing. \sys achieves this through a set of tightly integrated techniques: representative index caching (Sec.~\ref{sec:design:cache}), RDMA-friendly  graph index storage layout (Sec.~\ref{sec:design:layout}), query-aware batched data loading (Sec.~\ref{sec:design:load}), pipelined execution model (Sec.~\ref{sec:design:pipeline}) and index rebuild mechanism (Sec.~\ref{sec:design:rebuild}). Here we provide a brief overview of \sys as Fig.~\ref{fig:overview} shows. \sys consists of an offline part and a runtime part.}

\begin{figure}[!t]
	\centering
	\includegraphics[scale=0.4]{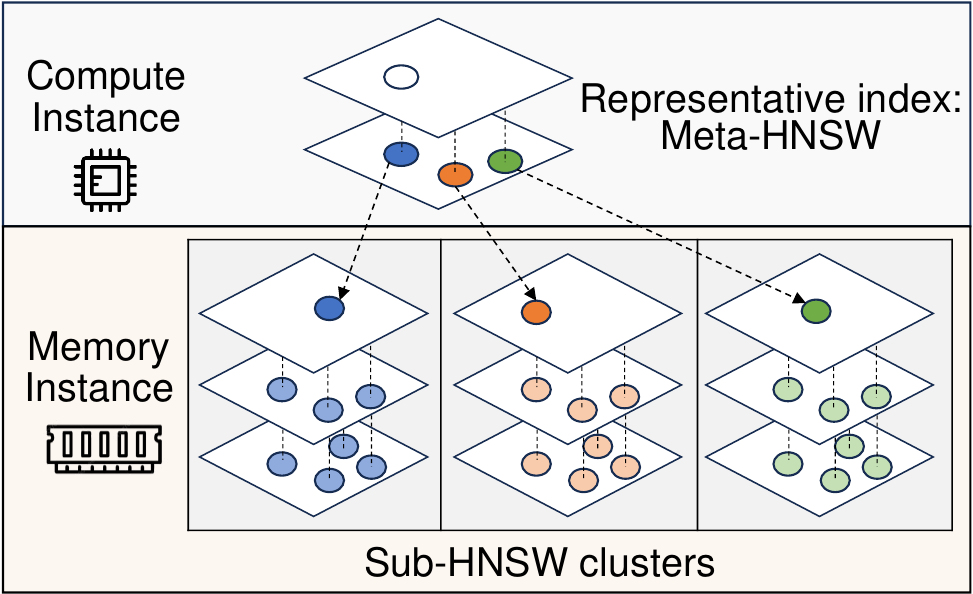}
    \vspace{-1ex}
	\caption{Meta-HNSW and Sub-HNSW.}
	\vspace{-1.5ex}
	\label{fig:represent}
    \vspace{-3.5ex}
\end{figure}
\textbf{Offline.} \sys first partitions the dataset into balanced subsets using an index balancer that applies K-Means++\cite{kmeans++} initialization with approximate sampling, followed by a priority queue-based assignment phase to enforce capacity constraints while minimizing intra-subset distances. This balanced partitioning ensures that each subset contains a uniform number of vectors, bringing predictable network transfer and uniform compute load across the distributed system. \rb{During offline construction, the system partitions the dataset into balanced
subsets and derives a centroid set for routing. The \emph{memory nodes} construct a lightweight representative index over these centroids, forming a \textit{meta-HNSW} that is stored in the disaggregated memory pool as query-time routing metadata, then builds \textit{sub-HNSWs} and materializes the index structures directly in disaggregated memory pool. \emph{Compute nodes} fetch metadata from memory nodes (e.g., the meta-HNSW, sub-HNSW offset ranges, and overflow descriptors) and cache this metadata for subsequent query routing and loading. Under dynamic workloads, as the reserved overflow gap is consumed by insertions, the system triggers \emph{rebuild} operations to rebalance partitions, preserve search quality, and support subsequent insertions.} \rc{Here, “overflow” denotes the excess vectors that surpass a sub-HNSW’s designated capacity due to workload dynamics or uneven data insertions. }Each compute instance also caches an offset table that records the start and end memory positions of each sub-HNSW index within its local address space. 

\textbf{Runtime.} At runtime, \sys processes vector queries that arrive at the system. \sys supports multiple distance metrics (e.g., Euclidean, angular) to accommodate diverse vector search scenarios. \sys has the following three major components in the query runtime. 

\textbf{1. Query routing.} After a query arrives at a compute instance, \sys performs a search on the cached meta-HNSW to identify the top-$b$ candidate sub-HNSWs that are most likely to contain relevant vectors for the query. This coarse-grained classification efficiently narrows down the search space to a small subset of relevant partitions without requiring network access. For a batch of B queries ${q_1, q_2, ..., q_B}$, each query requires searching the top-$k$ closest vectors from its $R$ closest sub-HNSWs, resulting in up to $R \times B$ required sub-HNSW accesses. 

\textbf{2. Query-aware loading.} The batch optimizer analyzes the required sub-HNSWs across all queries in the batch and ensures that each sub-HNSW is loaded from the memory pool only once, even if multiple queries require the same sub-HNSW cluster. Given that the DRAM resources in the compute instance can only accommodate and cache $c$ sub-HNSWs, the system schedules data transfers to minimize network overhead. The system also retains the most recently loaded $c$ sub-HNSWs for subsequent batches to further reduce data transfer overhead.

\textbf{3. Pipelined executor.} The pipelined executor orchestrates the vector search pipeline with network data transfer to optimize performance. It overlaps the data transfer of sub-HNSW clusters with the computation of both meta-HNSW and sub-HNSW search operations. When a sub-HNSW cluster is required, the compute instance issues an RDMA\_READ command to retrieve the cluster along with its corresponding shared overflow memory space, ensuring that newly inserted vectors are accessed contiguously with the original data in a single operation. The pipeline carefully manages scheduling to effectively hide the latency of transferring sub-HNSW data within the vector search time of the previous batch, thereby minimizing overall query latency. After finishing the computation of all required sub-HNSWs in a batch, the executor returns the final top-$k$ results to the client.

\rb{\textbf{4. Rebuild mechanism.}}
\rb{To sustain performance under dynamic insertions, \sys monitors the utilization of the overflow region and triggers a rebuild when the reserved capacity is exhausted. The rebuild is performed by constructing a shadow copy of the updated index in a new epoch buffer and atomically switching to the new epoch after serialization into a fresh RDMA region. During the transition, compute instances continue serving queries on the old epoch, while newly inserted vectors remain query-visible via a shared insertion buffer organized by LSH buckets. After the epoch switch, buffered inserts are merged into the rebuilt index, and compute instances refresh routing metadata and invalidate stale cached sub-HNSWs.}

\vspace{-1.ex}
\section{\sys Design}
\label{sec:design}
\re{In this section, we present the design of \sys that includes four main techniques. 1) \textit{Representative index caching} to enable more scalable and resource efficient designs (Sec.~\ref{sec:design:cache}). 2) \textit{RDMA-friendly graph index storage in remote memory} to support contiguous data placement and one-sided RDMA insertion with high throughput under insertions (Sec.~\ref{sec:design:layout}). 3) \textit{Query-aware batched data loading} to enable efficient batched operations (Sec.~\ref{sec:design:load}). 4) \textit{Orchestrating vector search and network data transfer pipeline} to reduce query latency by overlapping computation and data transfer (Sec.~\ref{sec:design:pipeline}).}

\subsection{Representative index caching}
\label{sec:design:cache}

\begin{figure}[!t]
	\centering
	\includegraphics[scale=0.46]{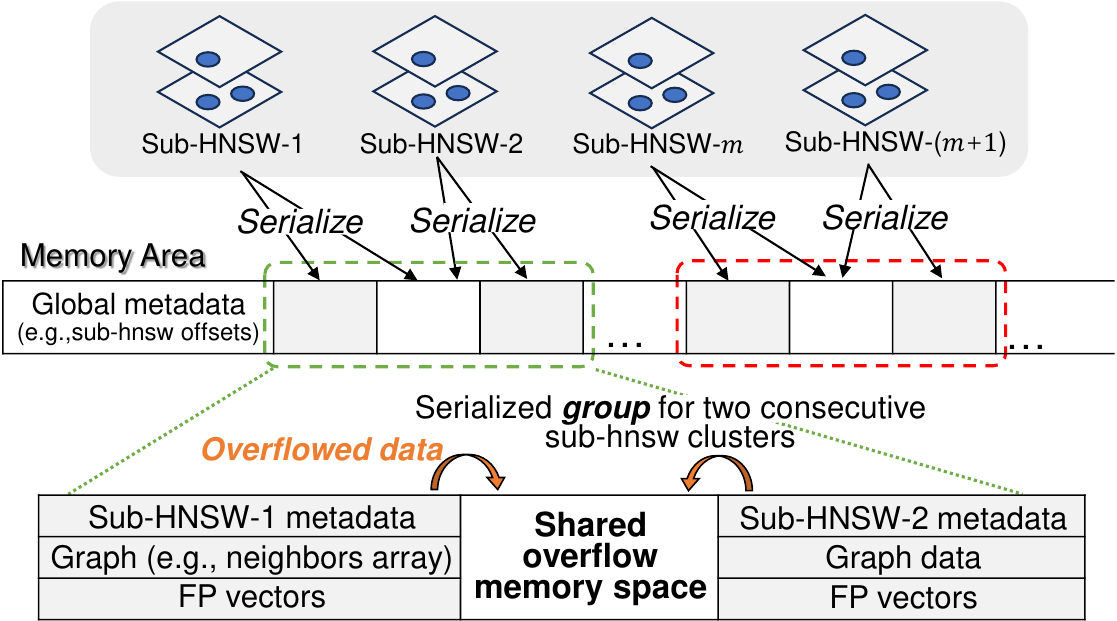}
    \vspace{-1ex}
	\caption{RDMA-friendly sub-HNSW indexing data layout in remote memory.}
	\vspace{-2.5ex}
	\label{fig:layout}
\end{figure}

Graph-based vector search schemes~\cite{nsg, hnsw} rely on greedy routing to iteratively navigate toward the queried vector. However, the search path can span the entire graph, potentially covering distant vectors. HNSW exhibits small-world properties, allowing long-range connections between vectors that are far apart. However, loading the entire graph index from the memory pool to the compute pool for each query is impractical because the compute pool has limited memory in a disaggregated system. This approach would not only consume excessive bandwidth by transferring a significant portion of untraversed vectors but also introduce additional latency, thereby degrading the overall search efficiency.
To enable efficient graph-based vector search in disaggregated memory environments, \sys adopts a \textbf{hierarchical indexing strategy} that partitions the dataset into balanced subsets while maintaining high recall and low query latency. This strategy comprises the following two key components: 

\subsubsection{Index Balancer}
\label{sec:design:cache:balancer}
We apply K-Means++~\cite{kmeans++} initialization with approximate sampling, followed by a priority queue-based assignment phase to enforce capacity constraints while minimizing intra-subset distances. This balanced partitioning ensures that each subset contains a uniform number of vectors, resulting in predictable network transfer and balanced per-partition computation, thereby preserving pipeline efficiency during large-scale query processing. Based on this partitioning, \sys derives a set of centroids that are later used for query routing.

\subsubsection{Representative Metadata}
\label{sec:design:cache:meta}
To efficiently route queries to the relevant partitions, as shown in Fig.~\ref{fig:represent}, we utilize a lightweight three-layer hierarchical HNSW graph over partition centroids from the sampled metadata introduced by Sec.~\ref{sec:design:cache:balancer}, referred to as \textit{meta-HNSW}. The meta-HNSW serves as representative metadata, enabling coarse-grained classification to efficiently identify a small subset of relevant partitions for a given query. We cache the meta-HNSW on every compute instance. At query time, the compute instance performs a search on the cached meta-HNSW to identify the top candidate sub-HNSWs, fetching only those sub-HNSWs from the memory pool via one-sided RDMA operations for subsequent search. 

\vspace{-1ex}
\subsection{RDMA-friendly graph index storage layout in remote memory}
\label{sec:design:layout}

\textbf{RDMA-Friendly Sub-HNSW Layout.}
RDMA enables efficient data access to targeted remote memory addresses. To efficiently read and write sub-HNSW cluster data in remote memory, an intuitive approach is to serialize all sub-HNSW clusters in the registered memory. 

When new vectors are inserted, the size of each sub-HNSW cluster may exceed the allocated space. However, shifting all stacked sub-HNSW clusters is impractical, newly inserted vectors in the same partition and their metadata may be placed in non-contiguous memory regions if they are simply appended at the tail of the available area. This fragmentation increases access latency and reduces query throughput due to the higher cost of scattered index access. In our design, we allocate and register a continuous memory space in the memory node to store both the serialized HNSW index and floating-point vectors, as shown in Fig.~\ref{fig:layout}. At the beginning of this memory space, a global metadata block records the offsets of each sub-HNSW cluster, as their sizes vary. The remaining memory space is divided into \textit{groups}, each of which is capable of holding two sub-HNSW clusters. Within each group, our RDMA-friendly layout incorporates two levels of reserved space to accommodate dynamic insertions. First, we introduce \textit{internal gaps} within individual sub-HNSW clusters between critical data structures—specifically between the levels array, offsets array, neighbors array, and floating-point vectors. These internal gaps enable in-place expansion of these arrays without requiring data reorganization. Second, between each pair of sub-HNSW clusters, we allocate a \textit{shared overflow memory gap} to accommodate larger insertions that exceed the internal gap capacity. The first section of each group stores the first serialized sub-HNSW cluster, which includes its metadata, neighbor array for HNSW, and the associated floating-point vectors. The second sub-HNSW cluster is placed at the end of the group. This dual-gap design ensures that newly inserted vectors and their metadata can be stored contiguously with the original sub-HNSW data, enabling RDMA access without fragmented remote reads.

\begin{wrapfigure}[10]{r}{0.5\textwidth}
  \hfill
  \vspace{-2ex}
  \includegraphics[width=\linewidth]{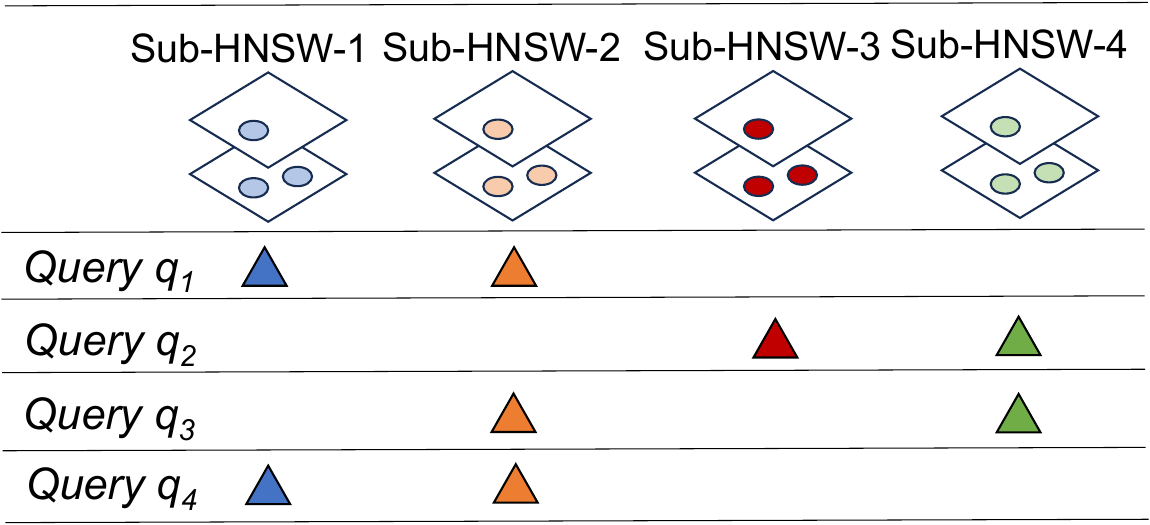}
  \vspace{-4.ex}
  \caption{\rd{Query-aware sub-HNSW clusters loading.}}
  \label{fig:load}
\end{wrapfigure}

\textbf{RDMA-Aware Operation Mechanism.}
When a vector query requires loading a sub-HNSW cluster, the compute instance issues an RDMA\_READ command to retrieve the cluster along with its corresponding shared overflow memory space. This layout ensures that newly inserted vectors are stored continuously with the original sub-HNSW data, enabling them to be read back with a one-time RDMA\_READ command. To optimize memory usage, each pair of adjacent sub-HNSW clusters shares a single overflow memory space for accommodating newly inserted vectors rather than allocating a separate one for each cluster. Also, the insertion mechanism leverages this RDMA-friendly layout for fast updates. Insertions are first routed to their target sub-HNSWs via the meta-HNSW. If the target sub-HNSW is already cached at the compute instance, the insertion is performed locally; otherwise, the target sub-HNSW is fetched from the memory pool and cached before applying the update. Rather than writing back the entire sub-HNSW, \sys tracks all modifications in an \textit{UpdateCommit}, which records metadata updates and data appends. \ra{The commit process utilizes \textit{doorbell batching} to write all changes, minimizing insertion latency. Doorbell batching reduces CPU overhead and PCIe signaling cost by allowing applications to post a batch of Work Queue Elements (WQEs) with a single doorbell write. This optimization requires only one memory-mapped I/O (MMIO) operation per batch, in contrast to the default WQE-by-MMIO approach, where each WQE incurs a separate MMIO, leading to significantly higher CPU and PCIe overhead.}
Newly inserted vectors are first placed using the reserved \textit{internal gaps} within each sub-HNSW cluster, enabling in-place growth of critical arrays without data reorganization. When an insertion exceeds the remaining internal gap capacity, the system appends the vector and its metadata sequentially to the \textit{shared overflow} region for that sub-HNSW group.
\rb{When any shared overflow region becomes full, the system triggers 
a rebuild, as detailed in Sec.~\ref{sec:design:rebuild}. This design avoids memory fragmentation and adapts to dynamic workloads, while maintaining low insertion overhead and preserving the efficiency of RDMA-based access.}

\vspace{-2.5ex}
\subsection{Query-aware batched data loading}
\label{sec:design:load}

\begin{figure}[!t]
	\centering
	\includegraphics[scale=0.38]{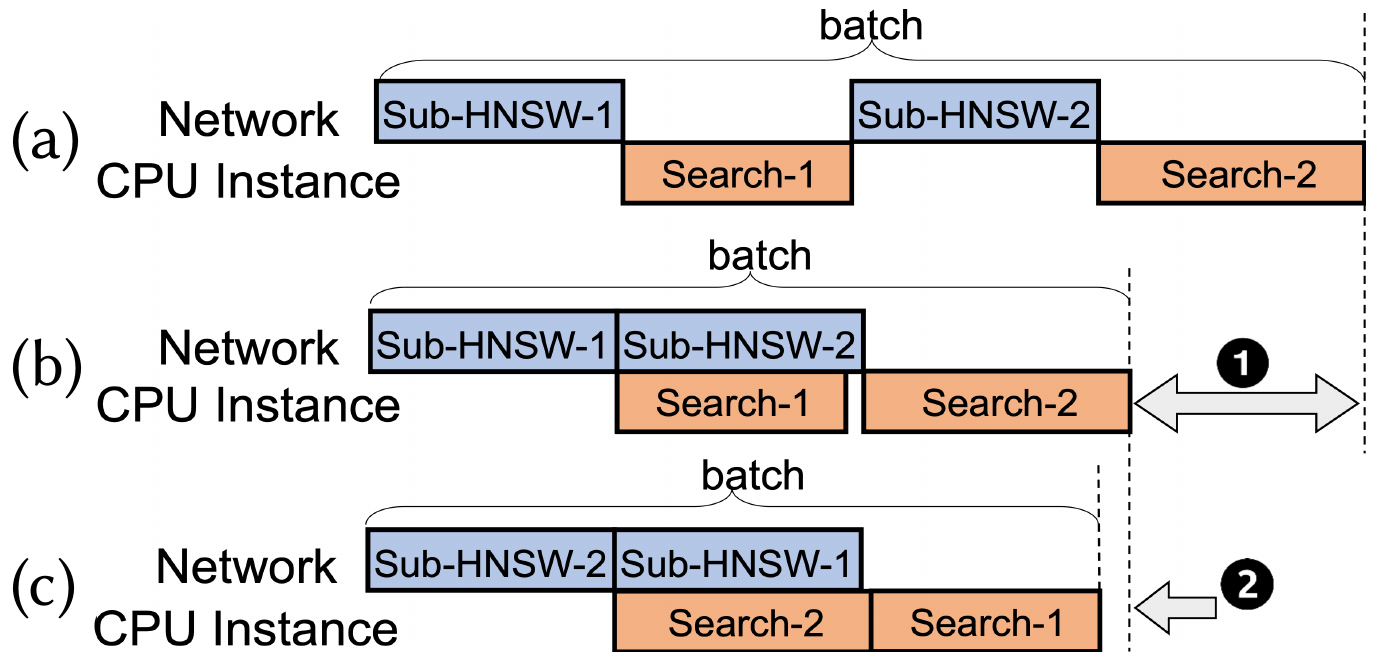}
    \vspace{-1ex}
	\caption{\rd{Orchestrating vector search and network data transfer pipeline.}}
	\vspace{-1.5ex}
	\label{fig:design:pipeline}
    \vspace{-2ex}
\end{figure}

To reduce network traffic for transferring graph index and improve query efficiency, we propose merging sub-HNSW index loading for queried vectors in the same batch.

Given a batch of queried vectors \{$q_1$,$q_2$, ...,$q_B$\} and a total of $P$ sub-HNSW clusters, each queried vector requires searching the top-$k$ closest vectors from the $R$ closest sub-HNSWs. However, the DRAM resources in the compute instance can only cache $c$ sub-HNSWs. To optimize loading, we analyze the required $R \times B$ sub-HNSWs accesses induced by the batch \textit{online} and ensure that each required sub-HNSW is loaded from the memory pool only \textbf{once} in one batch.

\re{For example, as shown in Fig.~\ref{fig:load}, queried vector \(q_1\) requires sub-HNSWs \(S_1\) and \(S_2\), while queried vector \(q_4\) searches the same two clusters. Similarly, query \(q_2\) and query \(q_4\) both require sub-HNSW \(S_2\). By identifying such overlaps across queries, the system avoids repeatedly loading the same sub-HNSWs within a batch. In the doorbell mode (see Sec.~\ref{sec:design:pipeline}), sub-HNSW loading is further batched at the RDMA level. For example, given a doorbell batch size of 2 for accessing sub-HNSWs, the compute instance can issue an RDMA\_READ command to fetch index $S_1$ and $S_2$ in one network round-trip, then compute the top-$k$ closest vectors candidates for queries \{$q_1$,$q_3$,$q_4$\} first.
\(q_1\) and \(q_4\) complete their search at this step, while query \(q_3\) requires an additional access to \(S_4\). \(q_2\), which is routed to \(S_3\) and \(S_4\), is handled in subsequent loading steps. Intermediate results are temporarily buffered and merged as additional sub-HNSWs are fetched to produce the final query results.}
\rb{To meet service-level objectives (SLOs), the system dynamically truncates batches in time. Specifically, if query arrivals are sparse or delayed, the doorbell batch is closed once the maximum waiting time threshold is reached, ensuring bounded query latency. }

To further optimize fetch scheduling, we observe that some sub-HNSW clusters are shared by multiple queries within the batch and thus have higher query loads. We prioritize these clusters for earlier loading to further reduce latency in Sec.~\ref{sec:design:pipeline}.

\rc{Once all required sub-HNSW clusters for the batched queried vectors have been loaded and traversed, the query results will be returned. Additionally, we employ an LRU-based local cache that retains the most recently loaded $c$ sub-HNSWs for the next batch. If the required sub-HNSWs are already in the compute instance, they do not need to be loaded again, further reducing data transfer overhead.}

\vspace{-2ex}
\subsection{Orchestrating vector search and network data transfer pipeline}
\label{sec:design:pipeline}
\rd{As shown in Fig.~\ref{fig:design:pipeline}(a), the original execution processes network data transfers and vector search sequentially, causing the CPU to stall while waiting for remote data and leading to inefficient query execution. To address this limitation, we propose a pipelined execution model that overlaps RDMA data transfers with vector search computation.}

To optimize overall performance, we orchestrate the vector search pipeline with the network data transfer. Specifically, we overlap the data transfer of the sub-HNSW index with the computation of both meta-HNSW and sub-HNSW search. As illustrated in Step~\fcirc{1} of Fig.~\ref{fig:design:pipeline}, by carefully managing the pipeline, the latency of transferring sub-HNSW data can be effectively hidden within the vector search time of the last batch


\rb{To further improve cache utilization and reduce cold-start latency, we reorder sub-HNSW fetches based on query demand,
where sub-HNSW clusters with higher query loads are prioritized for earlier fetching based on per-cluster query frequency statistics collected from cached meta-HNSW routing results. This design follows the reversed pipeline order shown in Step~\fcirc{2} of Fig.~\ref{fig:design:pipeline}. Since sub-HNSW clusters with more queries require longer search computation time, we leverage this extended computation time on previous clusters to overlap additional data transfers. }

The detailed pseudocode of the data query and insertion algorithms are provided in Appendix~\ref{sec:appendix:query_algo} and Appendix~\ref{sec:appendix:insert_algo}.

\subsection{\rb{Index Rebuild Mechanism}}
\label{sec:design:rebuild}
\begin{figure}[!t]
    \centering
    \begin{subfigure}[t]{0.475\linewidth}
        \centering
        \includegraphics[width=\linewidth]{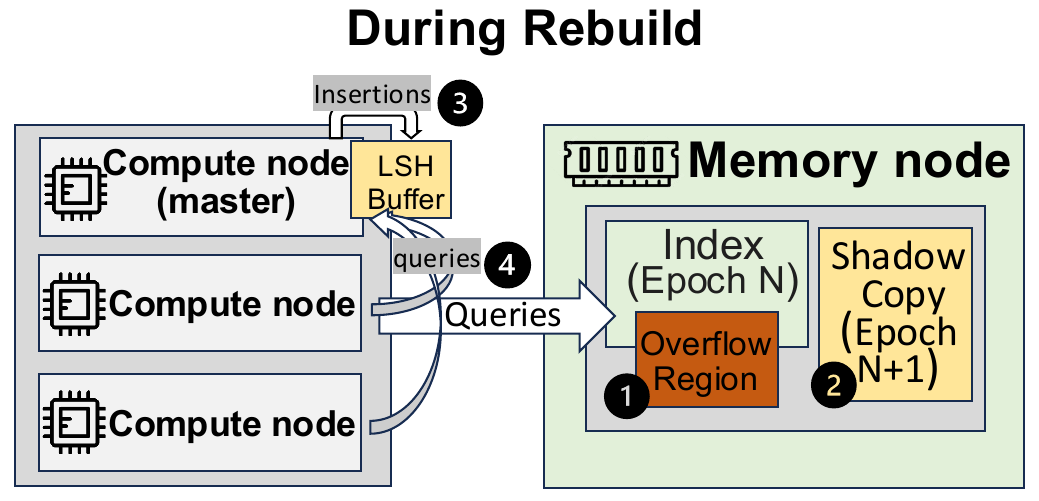}
        \vspace{-4.5ex}
        \caption{Serving queries and insertions during rebuild}
        \label{fig:design:rebuild:a}
    \end{subfigure}
    \begin{subfigure}[t]{0.475\linewidth}
        \centering
        \includegraphics[width=\linewidth]{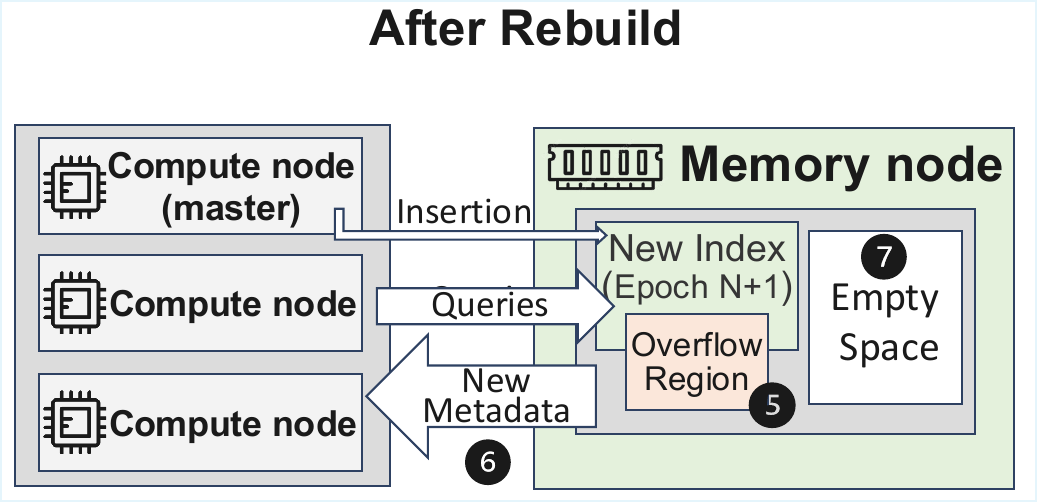}
        \vspace{-4.5ex}
        \caption{Switching epochs after rebuild}
        \label{fig:design:rebuild:b}
    \end{subfigure}
    \vspace{-2.5ex}
    \caption{\rb{Workflow of the index rebuild mechanism.}}
    \label{fig:design:rebuild}
    \vspace{-4ex}
\end{figure}
\rb{As \sys runs over time, dynamic insertions gradually consume the reserved overflow gap, eventually exhausting the available index capacity. In this situation, a rebuild is required to rebalance the index and restore its ability to accommodate further insertions. The key challenge is to perform such rebuilds while sustaining high query throughput, minimizing disruption to ongoing query processing, and preserving the visibility of newly inserted vectors. To address this challenge, \sys adopts an epoch-based shadow-copy rebuild mechanism that enables online index reconstruction atomic epoch transitions and ensures that newly inserted vectors remain visible to queries throughout the rebuild process.}

\rb{\sys monitors the utilization of overflow region and triggers a rebuild once the overflow region is full (Fig.~\ref{fig:design:rebuild:a} \fcirc{1}). Upon triggering, a designated master compute node responsible for insert coordination initiates a rebuild on the server side, which prepares a new epoch buffer and rebuilds the updated \sys index in a shadow copy (Fig.~\ref{fig:design:rebuild:a} \fcirc{2}), including updated sub-HNSW layouts and refreshed meta-HNSW routing metadata. The rebuilt state is serialized into a new RDMA region, after which the system commits the new epoch and safely transitions queries from the old epoch to the new one.}

\rb{During the transition, query processing proceeds uninterrupted on the old index. To preserve query freshness, newly inserted vectors are temporarily buffered in a shared auxiliary structure organized by LSH buckets (Fig.~\ref{fig:design:rebuild:a} \fcirc{3}), which is searched alongside the old-epoch sub-HNSWs during search (Fig.~\ref{fig:design:rebuild:a} \fcirc{4}). This ensures that recent inserts remain immediately query-visible even before the rebuild commits.}
\rb{After the rebuild completes, the system atomically commits the new epoch (Fig.~\ref{fig:design:rebuild:b} \fcirc{5}). Following the epoch switch, compute instances refresh routing metadata and invalidate stale cached sub-HNSWs (Fig.~\ref{fig:design:rebuild:b} \fcirc{6}). Once all queries have transitioned to the new epoch, the system safely deallocates the old index (Fig.~\ref{fig:design:rebuild:b} \fcirc{7}).}
\rb{Overall, this design enables online rebuilds that restore index balance and insertion capacity while preserving a continuous query service and consistent search results.}



\vspace{-1.5ex}
\subsection{\ra{Performance Modeling and Analysis}}
\label{subsec:modeling}

\ra{To better understand the performance characteristics of \sys, we develop analytical models for both the offline indexing stage and the online query serving stage.}

\begin{table}[t]
\centering
\caption{\ra{Key notations in the design.}}
\vspace{-2.3ex}
\renewcommand{\arraystretch}{0.92}
\footnotesize
\begin{tabular}{ll}
\toprule
\textbf{Symbol} & \textbf{Description} \\
\midrule
$N$ & Total number of database vectors. \\
$d$ & Dimensionality of each vector. \\
$P$ & Number of sub-HNSW. \\
$k$ & Top-$k$ result size for each query. \\
$e_{\text{build}}$ & HNSW construction candidate list size.\\
$e_{\text{meta}}$ & Search candidate list size for meta-HNSW.\\
$e_{\text{sub}}$ & Search candidate list size for sub-HNSW.\\
$B$ & Query batch size.\\
$S$ & Average size (in bytes) of a serialized sub-HNSW.\\
$P_{\text{fetch}}$ & Number of uncached sub-HNSWs fetched per batch.\\
$T_{\text{build}}$ & Total indexing time. \\
$T$ & Serving latency for one batch. \\
$T_{\text{meta}}$ & Meta-HNSW search time for one batch. \\
$T_{\text{net}}$ & Network transfer time for fetching sub-HNSWs. \\
$T_{\text{deser}}$ & Sub-HNSW deserialization time per batch. \\
$T_{\text{comp}}$ & Sub-HNSW search computation time per batch. \\
$W_{\text{net}}$ & RDMA network bandwidth.\\
\bottomrule
\end{tabular}
\label{tab:notations}
\vspace{-3.5ex}
\end{table}
\vspace{-1ex}
\subsubsection{\ra{Indexing Model}}

\ra{The \textbf{indexing model} captures the time to partition the dataset, build sub-HNSW graphs, and construct the meta-HNSW. The indexing phase consists of four stages.}

\ra{\textbf{Phase 1: Approximate K-Means++ Initialization.}
\sys uses an approximate K-Means++ sampling strategy to select initial cluster centers. For each of the $P$ centers, the algorithm samples a constant number of candidates (denoted $c_{\text{sample}}$) and selects the one maximizing the minimum distance to existing centers. The time complexity is:}
\ra{$T_{\text{init}} = O(P \cdot c_{\text{sample}} \cdot P \cdot d) = O(P^2 \cdot d)$}

\ra{\textbf{Phase 2: Capacity-Constrained K-Means Clustering.}
The clustering algorithm runs for a fixed number of iterations $I_{\text{max}}$ . Each iteration involves:}
\begin{itemize}
\vspace{-1ex}
\item \ra{Distance computation between all $N$ vectors and $P$ centroids: $O(N \cdot P \cdot d)$}
\item \ra{Maintaining the top-$L$ nearest centroids per vector using a heap: $O(N \cdot P \cdot \log L)$}
\item \ra{Capacity-constrained assignment via a global priority queue: $O(N \cdot \log N)$}
\item \ra{Centroid recomputation: $O(N \cdot d)$}
\vspace{-1ex}
\end{itemize}
\ra{Since $L$ is a small constant and $P \cdot d$ typically dominates $\log N$, the total clustering time is:
$T_{\text{cluster}} = O(I_{\text{max}} \cdot N \cdot P \cdot d) = O(N \cdot P \cdot d)$}

\ra{\textbf{Phase 3: Sub-HNSW Construction.}
Each of the $P$ sub-HNSWs is built independently over approximately $N/P$ vectors. Assuming balanced partitions, the HNSW construction complexity for a single sub-index is $O\left(\frac{N}{P} \cdot \log\frac{N}{P} \cdot e_{\text{build}} \cdot d\right)$. Since sub-index construction is in parallel, with $P$ threads, the elapsed time is:
$T_{\text{sub}} = O\left(\frac{N}{P} \cdot \log\frac{N}{P} \cdot e_{\text{build}} \cdot d\right)$}

\ra{\textit{(Assumption: Partitions are approximately balanced. The capacity-constrained clustering ensures each partition contains at most $\lceil N/P \rceil$ vectors. This assumption is validated in Sec.~\ref{eva:balance}.)}}

\ra{\textbf{Phase 4: Meta-HNSW Construction.}
The meta-HNSW is built over $P$ partition centroids. Computing centroids requires a single pass over all vectors: $O(N \cdot d)$. Constructing the meta-HNSW graph takes $O(P \cdot \log P \cdot e_{\text{build}} \cdot d)$. Thus:
$T_{\text{meta\_build}} = O(N \cdot d + P \cdot \log P \cdot e_{\text{build}} \cdot d)$}

\ra{\textbf{Total Indexing Time.}
Combining all phases:
$T_{\text{build}} = T_{\text{init}} + T_{\text{cluster}} + T_{\text{sub}} + T_{\text{meta\_build}}$
For typical configurations where $N \gg P$ and $P \cdot d \gg \log N$, the clustering phase and sub-HNSW construction dominate, yielding $T_{\text{build}} = O(N \cdot P \cdot d + \frac{N}{P} \cdot \log\frac{N}{P} \cdot e_{\text{build}} \cdot d)$.}

\subsubsection{\ra{Serving Model}}

\ra{The \textbf{serving model} characterizes per-batch latency for online query processing. The serving path consists of two main phases: meta-HNSW search followed by a three-stage pipelined sub-search.}

\ra{\textbf{Phase 1: Meta-HNSW Search.}
For a batch of $B$ queries, the meta-HNSW (containing $P$ centroid vectors) is searched to identify the top-R candidate partitions per query. Using parallelization across queries:
$T_{\text{meta}} = O\left(\frac{B}{n_{\text{threads}}} \cdot d \cdot e_{\text{meta}} \cdot \log P\right)$, where $n_{\text{threads}}$ is the number of threads for meta-search.}

\ra{\textbf{Phase 2: Pipelined Sub-Search.}
After meta-search identifies a set of candidate partitions, the system processes them through a three-stage pipeline with dedicated worker threads:}

\begin{itemize}[leftmargin=4ex,itemsep=0.ex,parsep=0.ex]
\vspace{-1ex}
\item \ra{\textbf{Fetch Stage} ($T_{\text{net}}$): Uncached sub-HNSWs are fetched from remote memory via RDMA\_READ. Let $P_{\text{fetch}}$ denote the number of unique sub-HNSWs that are not in the local LRU cache. Each fetch transfers $S$ bytes on average:
$T_{\text{net}} = \frac{P_{\text{fetch}} \cdot S}{W_{\text{net}}}$
\textit{(Note: $P_{\text{fetch}}$ depends on the cache hit ratio.)}}
\item \ra{\textbf{Deserialization Stage} ($T_{\text{deser}}$): Fetched raw bytes are parsed into in-memory HNSW index structures. The implementation uses a \texttt{DirectMemoryIOReader} that reads directly from the fetch buffer, reducing but not eliminating memory copies. The deserialization cost scales with the data size:
$T_{\text{deser}} = O(P_{\text{fetch}} \cdot S)$}
\item \ra{\textbf{Search Stage} ($T_{\text{comp}}$): HNSW search is executed on each required sub-index. Let $M$ denote the total number of (sub-index, query) pairs to process. Each sub-HNSW contains approximately $N/P$ vectors:
$T_{\text{comp}} = O\left(M \cdot d \cdot e_{\text{sub}} \cdot \log\frac{N}{P}\right)$
}
\vspace{-1.2ex}
\end{itemize}

\ra{\textbf{Pipeline Latency Analysis.}
The three stages operate concurrently via dedicated threads communicating through thread-safe queues. Let $t_{\text{fetch}}^{(i)}$, $t_{\text{deser}}^{(i)}$, and $t_{\text{search}}^{(i)}$ denote the per-task time for the $i$-th sub-index in each stage, where $i \in \{1, \ldots, P_{\text{fetch}}\}$.}

\ra{For a classic three-stage pipeline processing $P_{\text{fetch}}$ tasks, the total latency consists of:}
\begin{itemize}
\vspace{-1ex}
\item \ra{\textbf{Bottleneck stage}: The slowest stage determines steady-state throughput.}
\item \ra{\textbf{Fill latency}: Time for the first task to reach the final stage.}
\item \ra{\textbf{Drain latency}: Time for the last task to complete after entering the pipeline.}
\vspace{-1ex}
\end{itemize}

\ra{Assuming roughly uniform per-task times within each stage, let $\bar{t}_{\text{fetch}} = T_{\text{net}}/P_{\text{fetch}}$, $\bar{t}_{\text{deser}} = T_{\text{deser}}/P_{\text{fetch}}$, and $\bar{t}_{\text{search}} = T_{\text{comp}}/P_{\text{fetch}}$. The pipeline latency is:
$T_{\text{pipeline}} = \max\{T_{\text{net}}, T_{\text{deser}}, T_{\text{comp}}\} + \bar{t}_{\text{fill}} + \bar{t}_{\text{drain}}$
where $\bar{t}_{\text{fill}} = \bar{t}_{\text{fetch}} + \bar{t}_{\text{deser}}$ (first task traverses fetch and deserialize before search begins) and $\bar{t}_{\text{drain}} = \bar{t}_{\text{deser}} + \bar{t}_{\text{search}}$ (last task completes deserialize and search after final fetch). In practice, \sys further amortizes the fill and drain overheads via a fetch reordering policy (Sec.~\ref{sec:design:pipeline}), which prioritizes sub-HNSW clusters with higher query loads for earlier fetching.}

\ra{Simplifying to
$T_{\text{pipeline}} = \max\{T_{\text{net}}, T_{\text{deser}}, T_{\text{comp}}\} + \bar{t}_{\text{fetch}} + 2\bar{t}_{\text{deser}} + \bar{t}_{\text{search}}$,}
\ra{When $P_{\text{fetch}}$ is large, the fill/drain overhead becomes negligible relative to the bottleneck stage, and $T_{\text{pipeline}} \approx \max\{T_{\text{net}}, T_{\text{deser}}, T_{\text{comp}}\}$.}
\ra{Thus, the total query latency for processing a batch of $B$ queries is: $T = T_{\text{meta}} + T_{\text{pipeline}}$.}


\vspace{-.5ex}
\section{Evaluation}
\label{sec:eval}
\vspace{-.5ex}
In this section, we provide evaluation results on our prototype implementation of \sys. The results demonstrates its query latency vs recall (Sec.~\ref{eva:overall}), provides analysis on throughput scaling behavior (Sec.~\ref{eva:tput}) and latency under mixed search and insert workloads (Sec.~\ref{eva:insert}), conduct an ablation study of our design (Sec.~\ref{eva:ablation}), provide a detailed pipeline latency breakdown (Sec.~\ref{eva:pipeline}) and examine cache effectiveness (Sec.~\ref{eva:cache}) and evaluate balanced clustering strategy (Sec.~\ref{eva:balance}).

\vspace{-1ex}
\subsection{Methodology}
\textbf{Hardware and environments.}
We develop and evaluate the prototype of \sys on CloudLab~\cite{cloudlab} using real-world hardware. Our testbed consists of four Dell PowerEdge R650 servers, each equipped with two 36-core Intel Xeon Platinum CPUs, 256GB RAM, a 1.6TB NVMe SSD, and a Mellanox ConnectX-6 100Gb NIC. Three servers act as a compute pool, while one serves as the memory instance. We use up to \textbf{30 workers} from the compute pool and up to \textbf{300 sub-HNSW clusters} from the memory pool.

The system prototype of \sys with 18.4K lines of code in C++ is based on Faiss~\cite{faiss}, a state-of-the-art vector query processing system adopted by many vector databases.

\begin{table}[t]
    \centering
    \resizebox{0.82\linewidth}{!}{
    \renewcommand{\arraystretch}{0.9}
    \setlength{\tabcolsep}{2pt}
    \fontsize{5}{6}\selectfont
    \setlength{\aboverulesep}{0.2ex}
    \setlength{\belowrulesep}{0.2ex}
    \setlength{\defaultaddspace}{0.1ex}
    \setlength{\heavyrulewidth}{0.25pt} 
    \setlength{\lightrulewidth}{0.25pt} 
    \setlength{\cmidrulewidth}{0.25pt}  
    \begin{tabular}{ccccccc}
        \toprule
        \textbf{Dataset}  &  \textbf{Dim} & \textbf{\makecell{Base\\[-2.3pt]vectors}} & \textbf{\makecell{Query\\[-2.3pt]vectors}}  & 
        \textbf{Distance} & 
        \textbf{\makecell{Memory\\[-2.3pt]footprint}}&
        \textbf{\makecell{Meta-HNSW\\[-2.3pt]size}}\\
        \midrule
        TEXT10M & 200 & 10M & 10K & Angular & 47.3 GB & 227.8 KB\\
        SIFT10M & 128 & 10M & 10K & Euclidean & 32.4 GB & 171.6 KB\\
        DEEP10M & 96 & 10M & 10K & Euclidean & 26.4 GB & 146.6 KB\\
        GIST1M & 960 & 1M & 1K & Euclidean & 16.4 GB & 671.6 KB\\
        SIFT1M & 128 & 1M & 10K & Euclidean &  2.9 GB & 72.2 KB \\
        \bottomrule
    \end{tabular}
    }
    \vspace{1ex}
    \caption{Datasets benchmarked in Evaluation.}
    \label{tab:eval:dataset}
    \vspace{-6.5ex}
\end{table}

\textbf{Workloads.}
These five datasets are commonly used in previous ANN research~\cite{chen2021spann,peng2023iqan,su2024vexless,wang2021milvus}. TEXT10M~\cite{yandex2023bigann} consists of 10M 200-dimensional vectors under inner-product similarity. As a cross-modal dataset, it can be used to enhance the training and evaluation of multimodal LLMs.  In our prototype, it occupies 47.29 GB of memory, while the meta-HNSW index is lightweight at 227.8 KB, making it feasible to cache entirely on the client side. We also use SIFT10M~\cite{jegou2010product}, which is based on the Scale-Invariant Feature Transform (SIFT), a widely used computer vision algorithm for identifying and describing local features in images. We further evaluate DEEP10M~\cite{yandex2023bigann}, a 10 million vector dataset with 96 dimensions based on deep feature representations, using Euclidean distances. DEEP10M occupies 26.40 GB of memory with a compact meta-HNSW index of 146.6 KB. GIST1M~\cite{douze2009evaluation} contains 1 million GIST feature vectors that capture the global characteristics of images, including color, texture, and spatial structure. GIST1M poses a significant challenge for vector search algorithms due to its high dimensionality despite its smaller size.  In our prototype, it occupies 16.44 GB of memory, and its meta-HNSW index size is 671.6 KB. In addition, we include SIFT1M~\cite{jegou2010product}, a smaller dataset containing 1 million SIFT feature vectors that serves as a standard benchmark for assessing ANN system recall and latency at the million-scale with moderate dimensionality. It requires only 2.97 GB of memory, with a meta-HNSW index size of 72.2 KB.


\textbf{Baselines.}
To the best of our knowledge, there is no existing vector similarity search engine designed for disaggregated memory systems. \re{To enable comparison with \sys, we extend Pyramid~\cite{pyramid} to support disaggregated memory and implement a naive HNSW baseline using RPC over RDMA and include \texttt{distributed-faiss}, a gRPC-based distributed FAISS HNSW system that neither leverages RDMA nor operates under a disaggregated memory architecture.Specifically, we compare \sys to the following two RDMA-based baselines and one external distributed ANN baseline while keeping the same searching parameters under fine-tuned batch sizes:}

\begin{itemize}[leftmargin=4ex,itemsep=0.ex,parsep=0.ex]
\vspace{-.5ex}
\item \textbf{d-Pyramid}. We implement a version of Pyramid~\cite{pyramid}, a recent distributed HNSW system, in disaggregated memory and name it d-Pyramid. While Pyramid partitions the dataset into sub-datasets for index building and assigns queries to relevant partitions for efficient distributed processing, d-Pyramid extends this framework to operate over RDMA-based remote memory. Specifically, d-Pyramid stores Pyramid’s index structures in memory instances while caching the meta-HNSW graph on compute instances. Upon receiving a query, d-Pyramid first performs a search on the meta-HNSW to identify relevant partitions and then fetches the corresponding sub-datasets from the memory instances to perform searches for final query processing.

\item \textbf{\sys RPC}. To provide a straightforward baseline for comparison, we implement \sys RPC, a naive HNSW baseline over disaggregated memory using RPC over RDMA. In this design, the entire index is stored on remote memory instances, and compute instances issue search requests via RPC, leveraging RDMA to transfer data between compute and memory nodes. Upon receiving a query, the compute instance sends it to the memory instance, which performs the vector similarity search locally and returns the results to the compute instance. Unlike \sys, which only uses one-sided RDMA and requires no remote CPU involvement during query processing, \sys RPC incurs CPU overhead on memory instances for each query. 

\item \rc{\textbf{distributed-faiss (gRPC, non-disaggregated)}}. \re{We adopt the state-of-the-art distributed implementation of HNSW in \texttt{distributed-faiss}~\cite{DBLP:journals/corr/abs-2112-09924} as an additional gRPC-based baseline. The system shards the dataset across server processes, while each shard builds an independent FAISS HNSW index. Query requests are distributed to all shards through standard TCP-based RPCs, and a central coordinator merges the partial top-$k$ results to form the final answer. 
}
\end{itemize}
\vspace{-2.5ex}
\subsection{Overall Performance}
\vspace{-.5ex}
\label{eva:overall}
\begin{figure*}[t]
    \centering
    \includegraphics[width=1\textwidth]{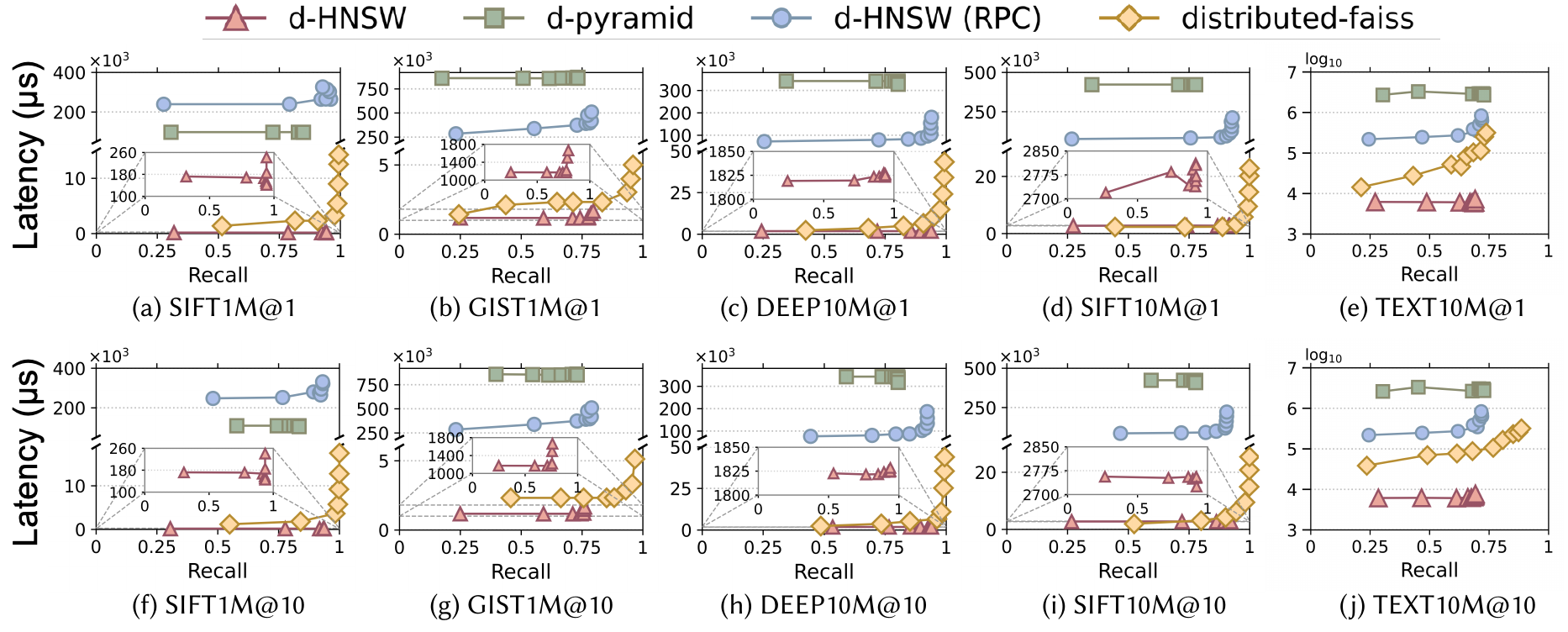}
    \vspace{-5.5ex}
    \caption{\re{Search Recall@1/10 latency Performance.}}
    \vspace{-4ex}
    \label{fig:lat_recall}
\end{figure*}

We evaluate the query latency and recall of \sys against d-Pyramid and \sys RPC across five types of vector datasets (1M to 10M vectors) under both top-1 and top-10 recall settings. 
For each dataset, we consider both top-1 and top-10 recall settings. We denote each workload as Dataset@k, where, for example, SIFT1M@1 indicates top-1 recall evaluation on the 1-million-scale subset of the SIFT dataset. To ensure fair SLO compliance, we gracefully tune the batch sizes for all baselines, maximizing their throughput while respecting the latency target. 

\re{As shown in Fig.~\ref{fig:lat_recall}, \sys consistently achieves \textbf{sub-millisecond to low-millisecond} query latency while preserving high recall across diverse workloads, representing \textbf{two to three orders of magnitude lower latency}
than existing disaggregated-memory baselines.}
Across all evaluated datasets, \sys completes queries within approximately \textbf{140--7{,}800~$\mu$s}, even at the 10M scale and under top-10 recall settings. By avoiding repeat partition loading and CPU bottlenecks, \sys significantly reduces both latency and system overhead. Through query-aware batched data loading and a pipelined execution model that overlaps RDMA transfers with computation, \sys approaches the practical lower bound of latency in disaggregated settings.
Compared to d-Pyramid, \sys shows 150--730$\times$ speedup across workloads. Frequent remote sub-dataset fetching in d-Pyramid causes significant network overhead. For instance, on SIFT1M@1, \sys achieves a latency of only 142.00~$\mu$s, compared to over 97.65~ms for d-Pyramid, resulting in a 656$\times$ improvement. Similarly, on high-dimensional workload GIST1M@1, \sys completes queries in 1172~$\mu$s versus 862.87~ms for d-Pyramid. Even on large-scale datasets like DEEP10M@10, \sys achieves 1824.21~$\mu$s, compared to 342.90~ms for d-Pyramid, corresponding to a 188$\times$ speedup.

\sys also significantly outperforms \sys RPC, achieving 40-2{,}100$\times$ lower latency. \sys RPC incurs high CPU load on memory nodes, leading to severe contention and overhead.  For example, on SIFT1M@1, \sys achieves a latency of just 142.00~$\mu$s compared to 265.76~ms for \sys RPC—a 1,871$\times$ improvement. Even on larger workloads like DEEP10M@10, \sys reduces query latency from 134.08~ms to 1,824.21~$\mu$s, achieving a 74$\times$ speedup.

Despite operating in a much lower latency regime, \sys provides comparable recall across all workloads across diverse scales and dimensionalities. \sys maintains similar or better recall compared to both baselines while achieving much lower latency. For example, in small-scale workload SIFT1M@1, compared to d-Pyramid, \sys achieves a significantly higher recall (94.24\% vs. 84.87\%) under the same number of partitions and identical $e_{search}$. This indicates that \sys's partitioning strategy more accurately routes queries to the correct sub-HNSW, enabling better candidate coverage. 
On high-dimensional workloads such as GIST1M@1, \sys achieves 79.63\% recall, which is above d-Pyramid’s 73.30\% and comparable to \sys RPC’s 78.92\%. Similar trends hold for large-scale workloads. On DEEP10M@10, \sys reaches a recall of 93.93\%, significantly outperforming d-Pyramid’s 80.30\% under identical partitioning and search settings, and closely matching \sys RPC’s 92.76\%.

 \re{Finally, \sys operates in a fundamentally lower latency regime compared with distributed-faiss, a distributed HNSW system using gRPC, achieving comparable recall with sub-millisecond to low-millisecond latency, whereas distributed-faiss requires tens to hundreds of milliseconds to approach high-recall operating points across workloads. For example, on SIFT1M@1, distributed-faiss achieves approximately 91\% recall at 2.32~ms, while further increasing recall (e.g., to 99.3\%) comes at the cost of substantially higher latency (over 14~ms). In contrast, \sys achieves comparable recall with sub-millisecond latency. On large-scale workloads such as TEXT10M@10, distributed-faiss requires over 200~ms to reach 88\% recall, whereas \sys achieves similar recall within 7--8~ms, resulting in more than an order-of-magnitude latency reduction. This trend persists across larger-scale and higher-dimensional datasets, highlighting the large latency gap between gRPC-based distributed ANN and RDMA-based disaggregated designs.}


\vspace{-1.5ex}
\subsection{Throughput with Varied Number of Workers}
\label{eva:tput}
\vspace{-.5ex}
\begin{figure*}[t]
\vspace{-1ex}
    \centering
    \includegraphics[width=1.03\textwidth]{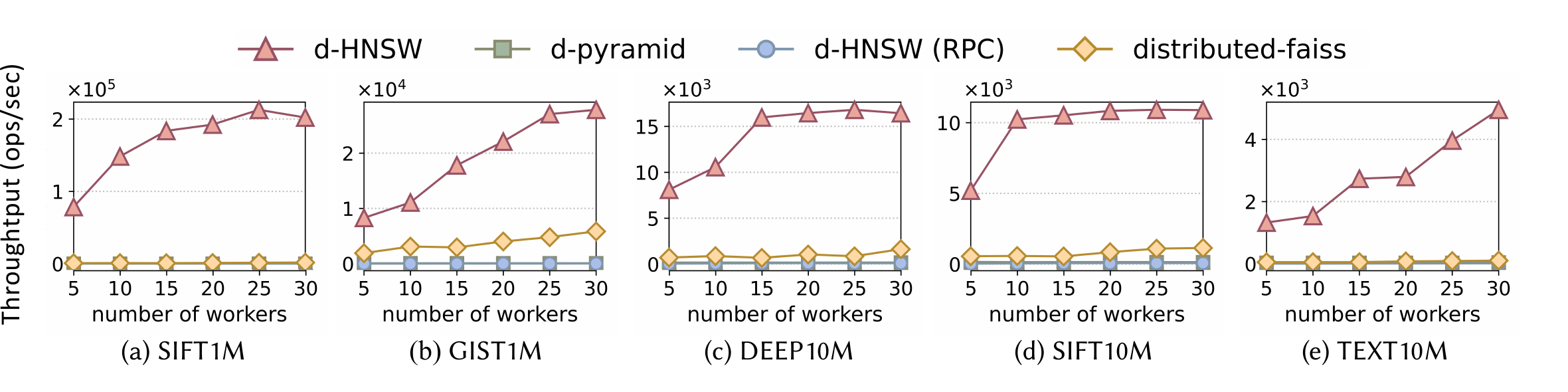}
    \vspace{-5.8ex}
    \caption{\re{Throughput under different worker counts.}}
    \label{fig:tput}
    \vspace{-3.5ex}
\end{figure*}

\begin{figure*}[t]
    \centering
    \begin{minipage}[t]{0.495\textwidth}
        \centering
        \begin{subfigure}[t]{0.49\linewidth}
            \centering
            \includegraphics[width=\linewidth]{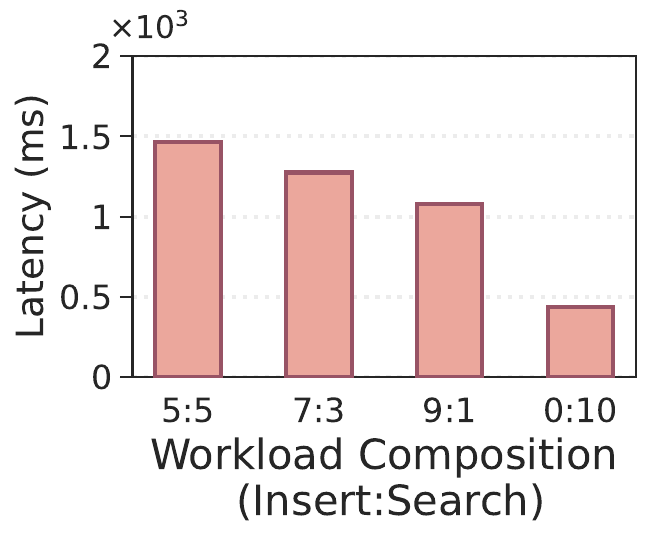}
            \vspace{-4ex}
            \caption{SIFT1M}
            \label{fig:insert:sift1M}
        \end{subfigure}\hfill%
        \begin{subfigure}[t]{0.49\linewidth}
            \centering
            \includegraphics[width=\linewidth]{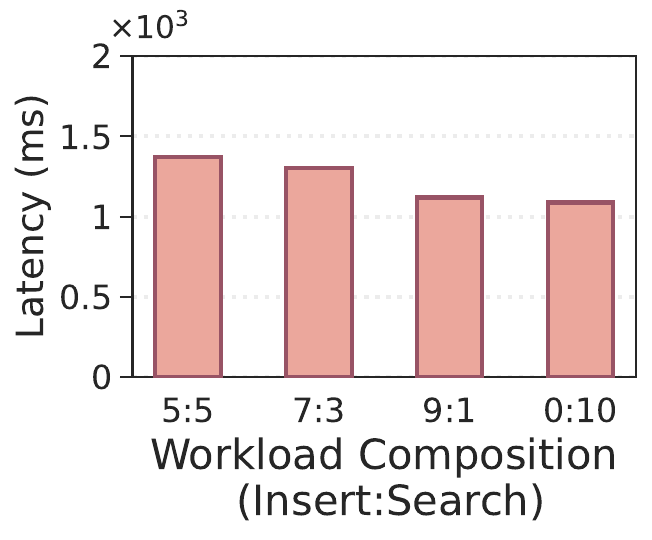}
            \vspace{-4ex}
            \caption{GIST1M}
            \label{fig:insert:gist1M}
        \end{subfigure}
        \vspace{-2.5ex}
        \caption{Operation batch latency under varying search-insert workload ratios.}
        \vspace{-1ex}
        \label{fig:insert}
    \end{minipage}\hfill%
    \raisebox{1.1ex}[0pt][0pt]{
    \begin{minipage}[t]{0.495\textwidth}
        \centering
        \begin{subfigure}[t]{0.49\linewidth}
            \centering
            \includegraphics[width=\linewidth]{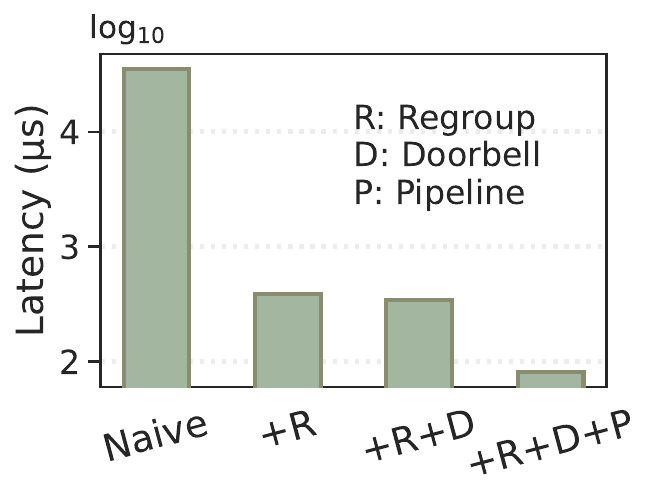}
            \vspace{-4ex}
            \caption{SIFT1M}
            \label{fig:ablation:sift1M}
        \end{subfigure}\hfill%
        \begin{subfigure}[t]{0.49\linewidth}
            \centering
            \includegraphics[width=\linewidth]{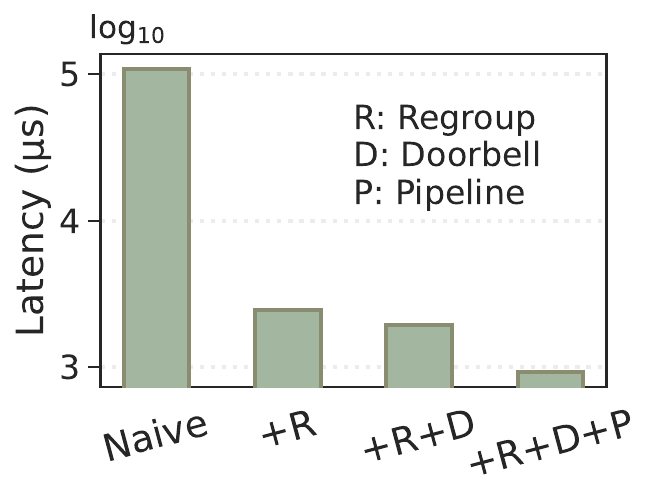}
            \vspace{-4ex}
            \caption{GIST1M}
            \label{fig:ablation:gist1M}
        \end{subfigure}
        \vspace{-2.5ex}
        \caption{Ablation study of search latency.}
        \vspace{-1ex}
        \label{fig:ablation}
    \end{minipage}}%
    \vspace{-4.ex}
\end{figure*}

We further evaluate how throughput of \sys and baselines scales with increasing numbers of compute workers. Specifically, we measure the system throughput under 5, 10, 15, 20, 25, and 30 workers across five workloads, as shown in Fig.~\ref{fig:tput}.

\sys demonstrates strong throughput scalability as the number of workers increases. On smaller workloads like SIFT1M and GIST1M, \sys demonstrates near-linear throughput growth up to 25 workers. For example, on SIFT1M, \sys reaches a throughput of 202K queries per second with 30 workers, which is 2.58$\times$ higher than the 5-worker configuration. On larger datasets such as SIFT10M and Deep10M, throughput improves significantly up to 10-15 workers and then with slower but steady gains beyond that point, reflecting increasing memory access overhead when querying large sub-HNSW clusters. Interestingly, on TEXT10M, \sys shows consistent throughput scaling, improving nearly 4$\times$ from 1,324 queries/s at 5 workers to 4,952 queries/s at 30 workers, suggesting better parallelism and lower compute node bottlenecks.

\re{In contrast, the baselines exhibit significantly weaker scalability. While d-Pyramid shows a little throughput improvement when increasing from 5 to 10 workers, the benefit quickly plateaus, and in cases like DEEP10M, adding more workers brings diminishing returns due to partition loading overhead and network contention. \sys RPC also scales poorly with additional compute workers, mainly constrained by the remote CPU bottleneck on memory nodes, leading to under-utilization of additional compute workers. Distributed-faiss achieves moderate throughput scaling as the number of workers increases, but remains fundamentally constrained by server-side CPU execution and TCP-based RPC communication. As a result, while distributed-faiss benefits from additional workers, its absolute throughput remains lower than \sys across all workloads.}

These results indicate that \sys effectively utilizes additional compute resources to scale throughput in disaggregated environments, even without scaling server-side network or bandwidth resources. Unlike d-Pyramid, \sys RPC, and distributed-faiss, which suffer from network bottlenecks or remote CPU overhead when adding more workers, \sys fully offloads query processing to compute nodes through one-sided RDMA operations. Notably, \sys achieves up to 124$\times$ higher throughput than baselines.

\vspace{-1ex}
\subsection{Latency under Mixed Search and Insert Workloads}
\label{eva:insert}

To demonstrate \sys’s support for dynamic insertions, we measure the latency of processing query batches under mixed workloads with different search-to-insert ratios. Fig.~\ref{fig:insert} reports batch processing time for fixed-size query batches (batch size = 8000) on SIFT1M and GIST1M datasets.

\sys supports concurrent search and insert operations. As the proportion of insert operations increases, batch latency grows due to resource contention from concurrent index updates. For example, on SIFT1M, batch latency increases from 438K~$\mu$s (100\% search) to 1.46M~$\mu$s (50\% insert), demonstrating that \sys remains functional under high insert rates while maintaining bounded latency.
These results demonstrate the ability of \sys to support runtime insertions along with query processing.

\vspace{-1ex}
\subsection{Ablation Study}
\label{eva:ablation}
\vspace{-.5ex}
To quantify the impact of each optimization in \sys, we perform an ablation study on SIFT1M and GIST1M datasets, measuring per-query latency while progressively enabling key system optimizations. Fig.~\ref{fig:ablation} presents the comparison of per-query latency under four configurations:
\begin{itemize}[leftmargin=3ex,itemsep=0.ex,parsep=0.ex]
    \vspace{-.5ex}
    \item \textbf{Naive}: a baseline disaggregated HNSW setup without any optimization;
    \item \textbf{+ Regroup}: adds query-aware batched data loading to minimize redundant memory accesses (see Sec.~\ref{sec:design:load});
    \item \textbf{+ Regroup + Doorbell}: incorporates RDMA doorbell batching to reduce network overhead;
    \item \textbf{+ Regroup + Doorbell + Pipeline}: replaces fixed doorbell batching with adaptive batching based on system load, additionally enabling pipelined execution to overlap computation and data transfer (see Sec.~\ref{sec:design:pipeline}).
    \vspace{-1.5ex}
\end{itemize}
The results show that each optimization yields additional latency reductions. On SIFT1M, per-query latency decreases from 34.8~ms in the naive design to 80.9~$\mu$s after applying all designs, achieving over $430\times$ improvement. Similarly, latency reduces from 107.9~ms to 924.5~$\mu$s with full optimizations, a $117\times$ improvement on GIST1M.


\begin{figure*}[t]
    \begin{minipage}[t]{0.51\textwidth}
        \vspace{-80pt}
        \begin{subfigure}[t]{0.49\linewidth}
            \centering
            \includegraphics[width=\linewidth]{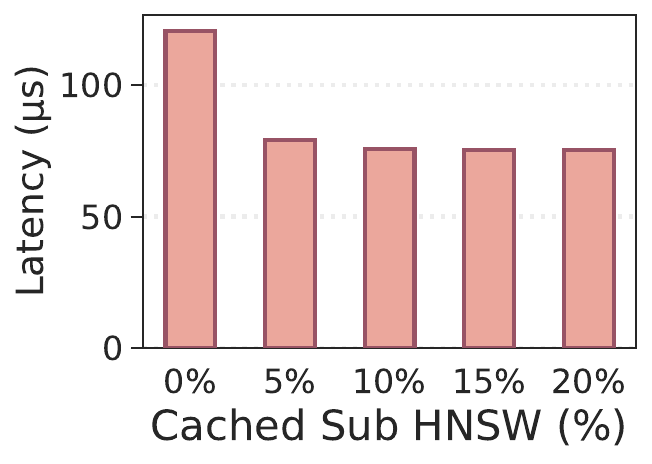}
            \vspace{-4.ex}
            \caption{SIFT1M}
            \label{fig:cache:sift1M}
        \end{subfigure}\hspace{-1ex}
        \begin{subfigure}[t]{0.49\linewidth}
            \includegraphics[width=\linewidth]{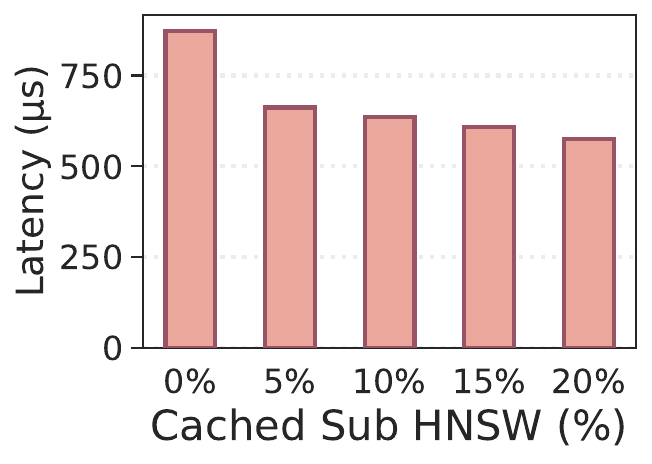}
            \vspace{-4.ex}
            \caption{GIST1M}
            \label{fig:cache:gist1M}
        \end{subfigure}
        \vspace{-2.5ex}
        \caption{Search latency under different cache ratios.}
        \vspace{-1ex}
        \label{fig:cache}
    \end{minipage}\hspace{-1ex}
    \begin{minipage}[t]{0.48\textwidth}
        \includegraphics[width=\linewidth]{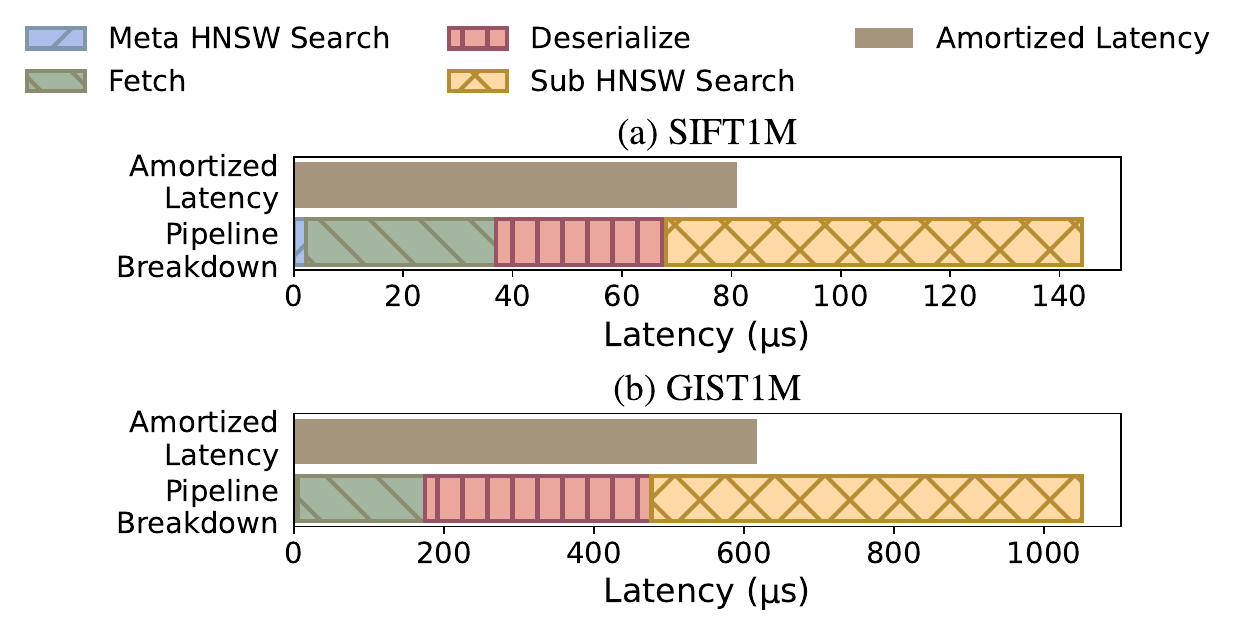}
        \vspace{-5ex}
        \caption{Pipeline latency breakdown.}
        \vspace{-1ex}
        \label{fig:pipeline}
    \end{minipage}
    \vspace{-2.4ex}
\end{figure*}

\vspace{-1ex}
\subsection{Query Latency under Different Cache Ratio}
\label{eva:cache}

We evaluate how the cache ratio of sub-HNSW affects query latency by varying the proportion of cached sub-HNSW from 0\% to 20\%. As shown in Fig.~\ref{fig:cache}, increasing the cache ratio consistently reduces query latency by reducing remote memory access frequency. For example, on SIFT1M, latency drops from 120.6~$\mu$s at 0\% cache to 75.6~$\mu$s at 10\%. Considering the trade-off between memory usage and query latency, we select a 10\% cache ratio as the default setting of \sys.

\vspace{-1ex}
\subsection{Pipeline Latency Breakdown}
\label{eva:pipeline}
\begin{figure*}[t]
    \centering
    \begin{minipage}[t]{0.57\textwidth}
        \centering
        \begin{subfigure}[t]{0.49\linewidth}
            \centering
            \includegraphics[width=\linewidth]{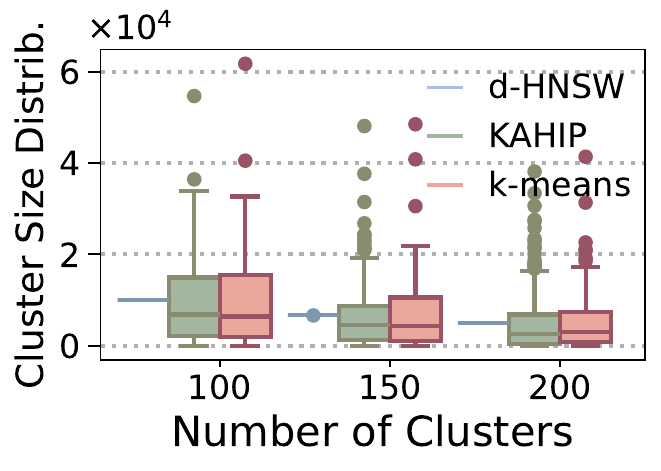}
            \vspace{-4ex}
            \caption{GIST1M}
            \label{fig:balance:gist1M}
        \end{subfigure}\hfill%
        \begin{subfigure}[t]{0.49\linewidth}
            \centering
            \includegraphics[width=\linewidth]{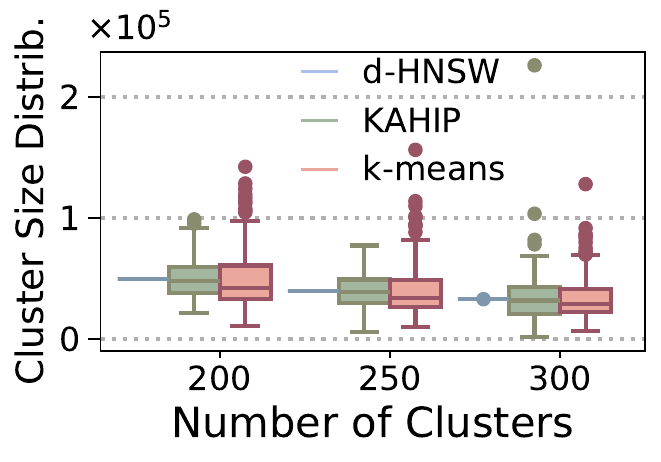}
            \vspace{-4ex}
            \caption{DEEP10M}
            \label{fig:balance:deep10M}
        \end{subfigure}
        \vspace{-2.8ex}
        \caption{Cluster size dist. with different partition rules.}
        \vspace{-1ex}
        \label{fig:balance}
    \end{minipage}\hfill
    \begin{minipage}[t]{0.385\textwidth}
        \centering
        \includegraphics[width=\linewidth]{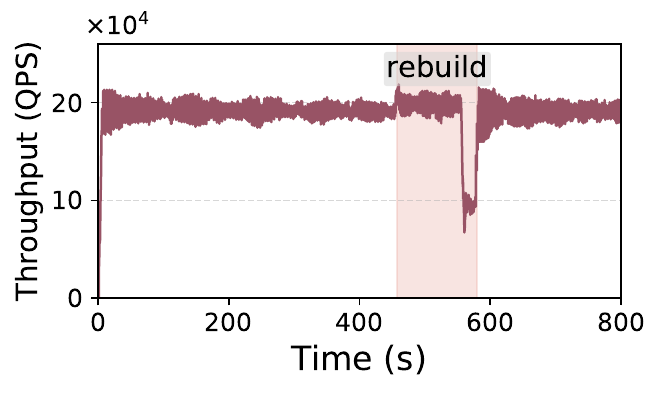}
        \vspace{-5ex}
        \caption{\rb{Throughput over time under index rebuilding.}}
        \vspace{-3ex}
        \label{fig:rebuild}
    \end{minipage}
    \vspace{-2ex}
\end{figure*}

{To validate the effectiveness of the pipelined execution, we break down the per-query latency into individual components: meta-HNSW search, RDMA fetch, deserialization, and sub-HNSW search. Fig.~\ref{fig:pipeline} reports the execution time of each component individually, as well as the query latency under pipelined execution.}

We observe that the total amortized latency is significantly smaller than the sum of the individual stage times, indicating that \sys successfully overlaps computation and data transfer through pipelined execution. For instance, on SIFT1M, while the sum of stage times exceeds 144~$\mu$s, the actual per-query latency is only 81.0~$\mu$s. Similarly, the total of all stages exceeds 1050~$\mu$s, the actual per-query latency is only 617.8~$\mu$s on GIST1M.

\vspace{-1ex}
\subsection{Balanced Partitioning Strategy}
\label{eva:balance}
\vspace{-.5ex}
As Fig.~\ref{fig:balance} shows, we analyze the cluster size distributions of \sys, KaHIP~\cite{sandersschulz2013} (used by d-Pyramid), and standard k-means~\cite{lloyd1982least} on GIST1M and DEEP10M datasets. KaHIP (Karlsruhe High Quality Partitioning) is a multilevel graph partitioning framework designed to minimize edge cuts while balancing partition sizes. \sys produces a near-balanced partitioning, with cluster sizes tightly concentrated and low variance. In contrast, both KaHIP and K-Means suffer from substantial size variance and and generate skewed clusters.


\vspace{-1ex}
\subsection{\rb{Rebuild}}
\label{sec:eval:rebuild}
\vspace{-.5ex}
\rb{We evaluate the d-HNSW’s ability to sustain throughput during online index rebuilding with one memory node, two search computing node (each running 10 workers), and one master computing node issuing a mixed workload of 20\% inserts and 80\% queries on the SIFT1M dataset.}

\rb{Fig.~\ref{fig:rebuild} plots the aggregated throughput of all three computing nodes over an 800-second interval, with the rebuild window (approximately 120 seconds) highlighted. Before rebuild, the system operates in a steady state, maintaining a stable throughput of approximately 190K queries per second under the mixed insert and query workload.}

\rb{Once rebuild is triggered (at $t \approx 480$s), the throughput briefly increases before exhibiting a significant drop. This transient increase happens because insert operations switch from issuing RDMA\_WRITE to the server to buffering locally at the insert client, thereby eliminating RDMA\_WRITE on the memory node. During this period, search workers continue serving queries using the old epoch while checking a small LSH buffer. The subsequent sharp throughput drop, falling to approximately 70K queries per second, coincides with the epoch transition. During this interval, search workers must suspend standard query processing to ensure safe handover between epochs. Specifically, workers stop issuing new search batches, wait for all in-flight RDMA operations referencing the old epoch to complete. Following this, an atomic metadata swap redirects pointers to the newly reconstructed index. After all clients acknowledge the new epoch, throughput rapidly returns to the level observed before the rebuild, indicating that the reconstructed index achieves comparable performance after the rebuild.}

\rb{Overall, the results show that \sys enables high-throughput query serving even during online index maintenance.}

%
%

%
%
%
%
%
%
%
%
%
%

\section{Related Work}
\label{sec:relatedwork}

\textbf{Disaggregated memory system.}
Disaggregated memory systems~\cite{shan2018legoos,wang2020semeru,memserve,mooncake,wang2024rcmp,cowbird,wei2023characterizing,wang2023disaggregated,wang2022case} have received increasing attention due to the ability to allocate flexible resources and improve hardware utilization in data centers. 
For example, recent works~\cite{zuo2022race,li2023rolex} addressed how to build high-performance key-value stores on disaggregated memory by exploiting one-sided RDMA operations to bypass remote CPU involvement.  
Outback~\cite{outback} proposes an RDMA RPC-based index by separating a dynamic minimal perfect hashing index into compute-heavy and memory-heavy components placed at compute and memory nodes. However, these systems cannot be applied to graph-based vector databases, as analyzed in Sec.~\ref{sec:background:challenges}.
\sys is orthogonal to these works.

\textbf{Approximate vector similarity search.}
Approximate similarity search has become a fundamental technique for efficiently retrieving high-dimensional data vectors in datasets. To balance accuracy and efficiency, prior work has developed a variety of indexing techniques, including tree-based structures (e.g., KD-trees~\cite{kdtree}), graph-based navigable small-world graphs (e.g., HNSW~\cite{hnsw}), and quantization-based approaches (e.g., product quantization~\cite{jegou2010product}). 

Among these, HNSW has emerged as a widely adopted method due to its excellent recall and low query latency. In contrast to prior works~\cite{li2020improving, zhang2019grip, hm-ann,auncel,su2024vexless,rummy,jang2023cxl} focused on monolithic or hardware-accelerated designs, \sys is the first to disaggregate the HNSW index into compute and memory pools over RDMA networks. \rb{Recent studies have also explored near-data-processing approaches using specialized hardware such as FPGAs or SmartNICs to execute ANN search directly within memory nodes~\cite{jiang2,isca}. While effective in reducing data movement, these designs require custom hardware and increase system cost and deployment complexity, whereas \sys achieves high performance using commodity RDMA NICs without introducing additional hardware dependencies.}

\section{Conclusion}
\label{sec:conclusion}
In this paper, we present \sys, the first RDMA-based vector similarity search engine specifically designed for disaggregated memory architectures. \sys addresses the fundamental challenges of deploying graph-based vector search algorithms like HNSW in disaggregated memory systems. \sys enhances vector request throughput and minimizes data transfer overhead through an RDMA-friendly data layout optimized for memory nodes. The system further optimizes batched vector queries by implementing query-aware data loading that eliminates redundant vector transfers, significantly reducing network bandwidth consumption and improving cache utilization. Through comprehensive evaluation across various datasets, we demonstrate that \sys outperforms other RDMA-based baselines by up to over $100\times$ in throughput.
These results validate the effectiveness of our co-designed approach that aligns vector search algorithms with the characteristics of RDMA-based disaggregated memory systems. 


\bibliographystyle{ACM-Reference-Format}
\bibliography{sample-base}

\appendix
\clearpage
\section{Appendix}



\subsection{Query Execution Algorithm}
\label{sec:appendix:query_algo}

The pseudocode of the query execution procedure and the fetch routine are outlined in Algorithm~\ref{alg:query} and Algorithm~\ref{alg:fetch}, respectively.
\begin{algorithm}[H]
\caption{Pipelined Query Execution Overview}
\label{alg:query}
\begin{algorithmic}[1]
\Require Query batch $Q[1..n]$
\Ensure Top-$k$ results: labels $L[1..n][1..k]$
\Statex \textit{//$L[i][j]$ is the label of the $j$-th nearest neighbor of query $Q[i]$}
\State $L_m \gets \Call{MetaSearch}{Q}$
\State Build query-to-subindex map $\mathcal{S}$ from $L_m$
\State $\mathcal{S} \gets \Call{SortByQueryCount}{\mathcal{S}, \texttt{desc}}$
\State Split $\mathcal{S}$ into:
\State $\mathcal{F}$: uncached shards
\State $\mathcal{R}$: cached shards
\State Initialize queues: \texttt{FetchQ} $\leftarrow \mathcal{F}$, \texttt{ReadyQ} $\leftarrow \mathcal{R}$

\State \textbf{/* The following are executed in parallel threads */}
\Statex \textbf{thread 1: Fetch Worker (persistent across batches)}
\While{\texttt{FetchQ} not empty}
    \State \Call{FetchWorker}{\texttt{FetchQ}, \texttt{DeserQ}}
\EndWhile

\Statex \textbf{Thread 2: Deserialize Worker (persistent across batches)}
\While{\texttt{DeserQ} not empty}
    \State $s \gets$ \Call{Dequeue}{\texttt{DeserQ}}
    \State $s \gets$ \Call{InPlaceDeserialize}{$s$}
    \Statex \textit{// reconstruct sub-HNSW $s$ directly from memory buffer}
    \State \texttt{DeserQ} $\gets$ \texttt{ReadyQ} $\cup$ \{ $s$ \}
\EndWhile

\Statex \textbf{Thread 3: Search Worker (persistent across batches)}
\While{\texttt{ReadyQ} not empty}
    \State $s \gets$ \Call{Dequeue}{\texttt{ReadyQ}}
    \State $L \gets$ \Call{SubHNSWSearch}{$s$, $Q_s$, $L$}
    \Statex \textit{// perform local search on sub-HNSW $s$ for query set $Q_s$ and merge top-$k$ results into $L$}
\EndWhile

\State \Return $L$
\end{algorithmic}
\end{algorithm}


\begin{algorithm}[t]
\caption{FetchWorker: Load sub-HNSW indices into DeserQ}
\label{alg:fetch}
\begin{algorithmic}[1]
\Require Fetch queue \texttt{FetchQ}, Deserialize queue \texttt{DeserQ}, Cached offset \texttt{Offset}, Cached shared overflow offset \texttt{Overflow}
\Statex \textit{//\texttt{Offset} records the start and end positions of each sub-HNSW in the memory pool. See Sec.~\ref{sec:design:overview} for details.}
\Statex \textit{// \texttt{Overflow} records prior overflow areas caused by in-place updates. See Sec.~\ref{sec:design:layout} for details.}
\If{doorbell mode enabled} 
\Statex  \textit{// doorbell mode details see Sec.~\ref{sec:design:pipeline}}
    \State Select up to $m$ sub-HNSW shards $s_1, \dots, s_m$ from \texttt{FetchQ}
    \Statex \textit{// $m$ is current cache capacity, typically smaller than the maximum doorbell batch size}
    \State $\texttt{DoorbellList} \gets \emptyset$
    \ForAll{$s_i$ in $\{s_1, \dots, s_m\}$}
        \State \texttt{DoorbellList} $\gets$ \texttt{DoorbellList} $\cup \{ \texttt{Offset}[s_i] \}$
        \If{\texttt{Overflow}[$s_i$] is defined}
            \State \texttt{DoorbellList} $\gets$ \texttt{DoorbellList} $\cup \{ \texttt{Overflow}[s_i] \}$
        \EndIf
    \EndFor
    \State \textsc{IssueDoorbellRDMA}(\texttt{DoorbellList})
    \If{\texttt{FetchStatus}[\texttt{DoorbellList}] = \texttt{Success}} 
        \State \texttt{DeserQ} $\gets$ \texttt{DeserQ} $\cup$ \{ $\{s_1, \dots, s_m\}$ \}
    \EndIf
\Else
    \State $s \gets$ \Call{Dequeue}{\texttt{FetchQ}}
    \If{\texttt{Overflow}[$s_i$] is defined}
        \State Issue doorbell RDMA command with \texttt{Offset}[$s$] and \texttt{Overflow}[$s$]
        \State \textsc{IssueDoorbellRDMA}({\texttt{Offset}[$s$], \texttt{Overflow}[$s$]})
    \Else
        \State \textsc{IssueRDMARead}( \texttt{Offset}[$s$])
    \EndIf
    \If{\texttt{FetchStatus}[$s$] = \texttt{Success}}
        \State \texttt{DeserQ} $\gets$ \texttt{DeserQ} $\cup$ \{ $s$ \}
    \EndIf
\EndIf
\end{algorithmic}
\end{algorithm}

\subsection{Insert Execution Algorithm}
\label{sec:appendix:insert_algo}
The pseudocode of the query execution procedure and the commit for update are outlined in Algorithm~\ref{alg:insert} and Algorithm~\ref{alg:update}, respectively.

\begin{algorithm}[t]
\caption{Insertion Execution Overview}
\label{alg:insert}
\begin{algorithmic}[1]
\Require Insert batch $I[1..n]$, Cache capacity $c$
\Ensure  Insertion status in remote memory pool
\State $L_m \gets \Call{MetaSearch}{Q}$
\State Build query-to-subindex map $\mathcal{S}$ from $L_m$
\State Split $\mathcal{S}$ into:
\State $\mathcal{C}$: cached shards
\State $\mathcal{U}$: uncached shards
\Statex \textbf{Process cached sub-HNSW:}
\ForAll{$i \in \mathcal{C}$}
    \State $\text{sub\_index} \gets \text{cache}[s]$
    \State \Call{PrepareAndCommitUpdate}{$i$, $\mathcal{S}[i]$}
\EndFor

\Statex \textbf{Process uncached sub-HNSW:}
\ForAll{$i \in \mathcal{U}$}
    \State $\text{sub\_index} \gets$ \Call{FetchSubHNSW}{$i$}
    \State $\text{cache}[s] \gets \text{sub\_index}$
    \State $\text{cache\_lru}.\Call{push\_front}{i}$
    \If{$|\text{cache}| > c$}
        \State \Call{EvictFromCache}{ }
    \EndIf
    \State \Call{PrepareAndCommitUpdate}{$i$, $\mathcal{S}[i]$}
\EndFor

\end{algorithmic}
\end{algorithm}

\begin{algorithm}[t]
\caption{PrepareAndCommitUpdate: Update sub-HNSW and commit via RDMA}
\label{alg:update}
\begin{algorithmic}[1]
\Require sub-HNSW $i$, Insertion data $\mathcal{S}[i]$, Cached Offset \texttt{Offset}, Cached shared overflow offset \texttt{Overflow}
\Ensure Successful update and RDMA commit of sub-HNSW $i$
\Statex \textit{// Extract current metadata from $i$}
\State Define $M = \{ntotal, entry\_point, max\_level, neighbors\_size, levels\_size, offsets, xb\_sizes\}$
\State Define $G = \{\texttt{levels},\ \texttt{offsets},\ \texttt{xb} \}$ \textit{// Extract current data structure from $i$, except for neighbors}
\State $I.\texttt{add}(\mathcal{S}[i])$ \textit{// Perform local insertion}
\Statex \textit{// Prepare commit record}
\State Initialize commit record: $\mathcal{C} \gets \{\text{.sub\_index} = i\}$
\State Identify modified metadata fields: $\Delta M \gets \{m \in M \mid I.m \neq M.m\}$
\ForAll{$m' \in \Delta M$}
\State $\mathcal{C} \gets \{\texttt{.overwrite} = \{m', \texttt{Offset[m]}\},sizeof(m')\}$
\EndFor
\State Identify data structure growth for $levels, offsets, xb$ with overflow management: $\Delta G \gets \{g \in G \mid \text{sizeof}(s.g) > \text{sizeof}(G.g)\}$
\ForAll{$g' \in \Delta G$}
\State $\mathcal{C} \gets \{\text{.overwrite} = \{m', \texttt{Offset[i][m]}\},sizeof(m')\}$
\If{internal gap size in \texttt{Offset}[i][g'] $\geq$ required size}
        \State Append write to $\mathcal{C}.\texttt{append} \gets \{g', \texttt{Offset}[i][g'], sizeof(g')\}$  
    \ElsIf{internal gap + overflow gap $\geq$ required size}
        \State Fill available space in \texttt{Offset}[i][g'] to $\mathcal{C}.\texttt{append}$ 
        \If{$i$ is the first sub-HNSW in its pair(details in Sec.~\ref{sec:design:layout})}
            \State  remainder in \texttt{Overflow}[i][g'] append regions to $\mathcal{C}.\texttt{append}$ in forward direction
        \Else
             \State  remainder in \texttt{Overflow}[i][g'] append regions to $\mathcal{C}.\texttt{append}$ in backward direction
        \EndIf
        \State Append both regions to $\mathcal{C}.\texttt{append}$ 
    \Else
        \State Trigger reconstruction
    \EndIf
\EndFor
\State Identify dirty regions for $neighbors$: $\Delta N \gets \{[s, e] \mid I.neighbors[s:e] \neq M.neighbors[s:e]\}$
\ForAll{region $[s, e] \in \Delta N$}
    \If{$e < n_{\text{neighbors}}$}
        \State Append write to $\mathcal{C}.\texttt{overwrites} \gets \{s, \texttt{Offset[i][neighbors]}\},e-s\}$
    \Else
    \State Handle append-only growth for $neighbors$ similar to $levels, offsets, xb$ (see lines~12--23)
    \EndIf
\EndFor
\Statex \textit{// Execute RDMA batch commit}
\State Prepare RDMA operations: $\text{ops} \gets \text{BuildRDMAOps}(\mathcal{C})$
\State \Call{ExecuteRDMABatch}{\text{ops}}
\State Update \texttt{Overflow}
\end{algorithmic}
\end{algorithm}
\end{document}